\documentclass[10pt,aps,prb,groupedaddress,twocolumn,amsmath,amssymb,eqsecnum,tighten,subentry]{revtex4-1}

\usepackage{nicefrac,bm,mathtools}                      
\usepackage{graphicx}                                   
\usepackage[pdftex]{hyperref}                           
\usepackage{siunitx}
\usepackage{xr}
\usepackage{verbatim}
\usepackage[version=4]{mhchem}

\usepackage{xcolor}

\newcommand{\figref}[2][]{Fig.~\ref{fig:#2}#1}
\newcommand{\secref}[1]{Sec.~\ref{sec:#1}}
\newcommand{\appref}[1]{App.~\ref{sec:#1}}
\newcommand{\eqnref}[1]{Eq.~(\ref{eq:#1})}

\newcommand{\subfig}[1]{\textbf{(#1)}}

\newcommand{\arr}[2]{\begin{array}{#1}#2\end{array}}
\newcommand{\dd}[1]{\mathrm{d}#1}
\newcommand{\ddd}[2]{\ensuremath{\operatorname{d}^{#2}\!{#1}}}
\newcommand{\pdb}[2]{\frac{\partial #1}{\partial #2}}
\newcommand{\punc}[1]{\,#1}

\newcommand{\ket}[1]{\left|#1\right\rangle}
\newcommand{\bra}[1]{\left\langle#1\right|}

\newcommand{\aR}{\alpha_{\mathrm{R}}}
\newcommand{\kR}{k_{\mathrm{R}}}
\newcommand{\ER}{E_{\mathrm{R}}}
\newcommand{\eF}{\epsilon_{\mathrm{F}}}
\newcommand{\eFpm}{\epsilon^\pm_\mathrm{F}}
\newcommand{\eFeq}{\epsilon^{\mathrm{eq.}}_\mathrm{F}}
\newcommand{\ek}{\epsilon_k}
\newcommand{\ekf}{\epsilon_k^{\mathrm{f}}}

\newcommand{\ekfFpm}{\epsilon_{\kF^\pm}^{\mathrm{f}}}
\newcommand{\evk}{\epsilon_{\mathbf{k}}}
\newcommand{\evkp}{\epsilon_{\mathbf{k}'}}

\newcommand{\vk}{\mathbf{k}}
\newcommand{\rso}{\mathbf{r}_{\mathrm{SO}}}
\newcommand{\muB}{\mu_{\mathrm{B}}}
\newcommand{\kB}{k_{\mathrm{B}}}

\newcommand{\cph}{c_{\mathrm{ph}}}

\newcommand{\mueq}{\mu^{\mathrm{eq.}}}
\newcommand{\dmu}{\delta\mu}
\newcommand{\kS}{k_{\mathrm{S}}}
\newcommand{\kF}{k_{\mathrm{F}}}

\newcommand{\kFO}{\kF^0}

\newcommand{\kk}[1]{\mathbf{k}_{#1}}
\newcommand{\rr}[1]{\mathbf{r}_{#1}}
\newcommand{\e }[1]{\epsilon_{#1}}
\newcommand{\ee}[1]{\epsilon_{\kk{#1}}}
\newcommand{\f}[1]{f(#1)}
\newcommand{\ff}[1]{f_{#1}}
\newcommand{\fermi}[2]{f_{#1}(#2)}
\newcommand{\qq}{\mathbf{q}}
\newcommand{\pp}{\mathbf{p}}

\newcommand{\vol}{\mathcal{V}}
\newcommand{\Ceq}{C_{\mathrm{eq.}}}
\newcommand{\dCbydt}{\frac{\dd{C}}{\dd{t}}}

\newcommand{\expp}[1]{\mathrm{e}^{#1}}
\newcommand{\I}{\mathrm{i}}

\newcommand*{\tran}{^{\mkern-1.5mu\mathsf{T}}}

\begin{document}

\title{Chirality relaxation in low-temperature strongly Rashba-coupled systems}

\author{P. C. Verpoort, V. Narayan}
\affiliation{
  Department of Physics, University of Cambridge, J.J. Thomson Avenue, Cambridge CB3 0HE, UK. \\
	Correspondence to: P. C. Verpoort $\langle$\href{mailto:pcv22@cam.ac.uk}{\nolinkurl{pcv22@cam.ac.uk}}$\rangle$, and V. Narayan $\langle$\href{mailto:vn237@cam.ac.uk}{\nolinkurl{vn237@cam.ac.uk}}$\rangle$}

\date{\today}

\begin{abstract}
We study the relaxation dynamics of non-equilibrium chirality distributions of charge carriers in Rashba systems. We find that at low temperature inter-Rashba band transitions become suppressed due to the combined effect of the Rashba momentum split and the chiral spin texture of a Rashba system. Specifically, we show that momentum exchange between carriers and the phonon bath is effectively absent at temperatures where the momentum of thermal phonons is less than twice the Rashba momentum. This allows us to identify inter-carrier scattering as the dominant process by which non-equilibrium chirality distributions relax. We show that the magnitude of inter-carrier scattering is strongly influenced by the opposing spin structure of the Rashba bands. Finally, we provide an explicit result for the inter-band relaxation timescale associated with inter-carrier Coulomb scattering. We develop a general framework and assess its implications for GeTe, a bulk Rashba semiconductor with a strong Rashba momentum split.
\end{abstract}


\maketitle

\section{Introduction}

The generation and control of electronic non-equilibrium states is of great importance for both understanding condensed matter systems, as well as for the design of new devices relevant for technological applications. Its feasibility hinges on a long relaxation timescale of those non-equilibrium states, which has sparked great interest in the search for long-lived electronic states in solid-state systems over the last few decades. For example, the endeavour to replace semiconductor-based technologies by spintronics devices has led to major interest in spin-relaxation timescales in spin-polarised systems~\cite{Wolf01}. Electronic momentum relaxes typically on timescales on the order of a picosecond, while spin-relaxation timescales can range up to microseconds at the longest~\cite{wu_spin_2010,zutic_spintronics_2004}. In this manuscript, we study the non-equilibrium relaxation timescales of yet another property of electronic charge carriers: chirality.

The chirality of charge carriers can only be defined well in systems that break spatial inversion symmetry, and we will focus our discussion on systems with strong Rashba spin-orbit coupling~\cite{Bihlmayer15}. Being a combination of both spin and momentum locked together, we will see how, at low temperatures, the chirality is protected from scattering events that in non-Rashba materials cause fast relaxation into equilibrium, and consequently we expect the chirality to exhibit long relaxation timescales. We will argue that this occurs due to a mismatch of the thermal phonon and the Rashba energy-momentum scales at low temperatures. The resulting dominant relaxation mechanism, which is the inter-carrier Coulomb interaction, is further weakened by the helical spin structure that is induced by the Rashba coupling.

Rashba systems are a technologically-relevant class of systems in which spatial inversion symmetry is absent and strong spin-orbit coupling causes the energy dispersion to be non-degenerate in spin. There is, therefore, much interest in Rashba systems as a potential platform for spintronics~\cite{Zutic04}, and towards manipulating spin currents using external fields~\cite{Nitta97}, and artificially-engineered heterostructures~\cite{Miller03, Eschbach15, Nguyen16, Backes17, Backes19}. Consequently, understanding non-equilibrium dynamics in those systems is of great relevance.

The focus of this contribution is to study the relaxation of chirality in Rashba systems in general, however we analyse the implications of our findings in the ferroelectric bulk-Rashba semiconductor GeTe, which ranks amongst the systems with the highest observed Rashba coupling and Rashba momentum split~\cite{Picozzi14}. We take various expressions for a general Rashba system obtained throughout this work, and evaluate them for the GeTe system.

This contribution is organised as follows. In \secref{model}, we provide an introduction to the theoretical framework used, as well as a general explanation of the non-equilibrium chirality distribution we intend to study. Next in \secref{phonons}, we study the relaxation of this state via phonon scattering, and in \secref{coulomb} we deal with relaxation through the Coulomb interaction. Finally, we conclude in \secref{conclusions}. Appendices \ref{sec:excess} and \ref{sec:app-time-constants} provide, respectively, explicit calculations of maximum allowed non-equilibrium carrier occupations and the inter-carrier relaxation time constants, while \appref{spinflips} explains the role of spin flips in the context of this work.

\section{Theoretical Framework}
\label{sec:model}

\begin{figure*}[t]
  \includegraphics[width= \textwidth]{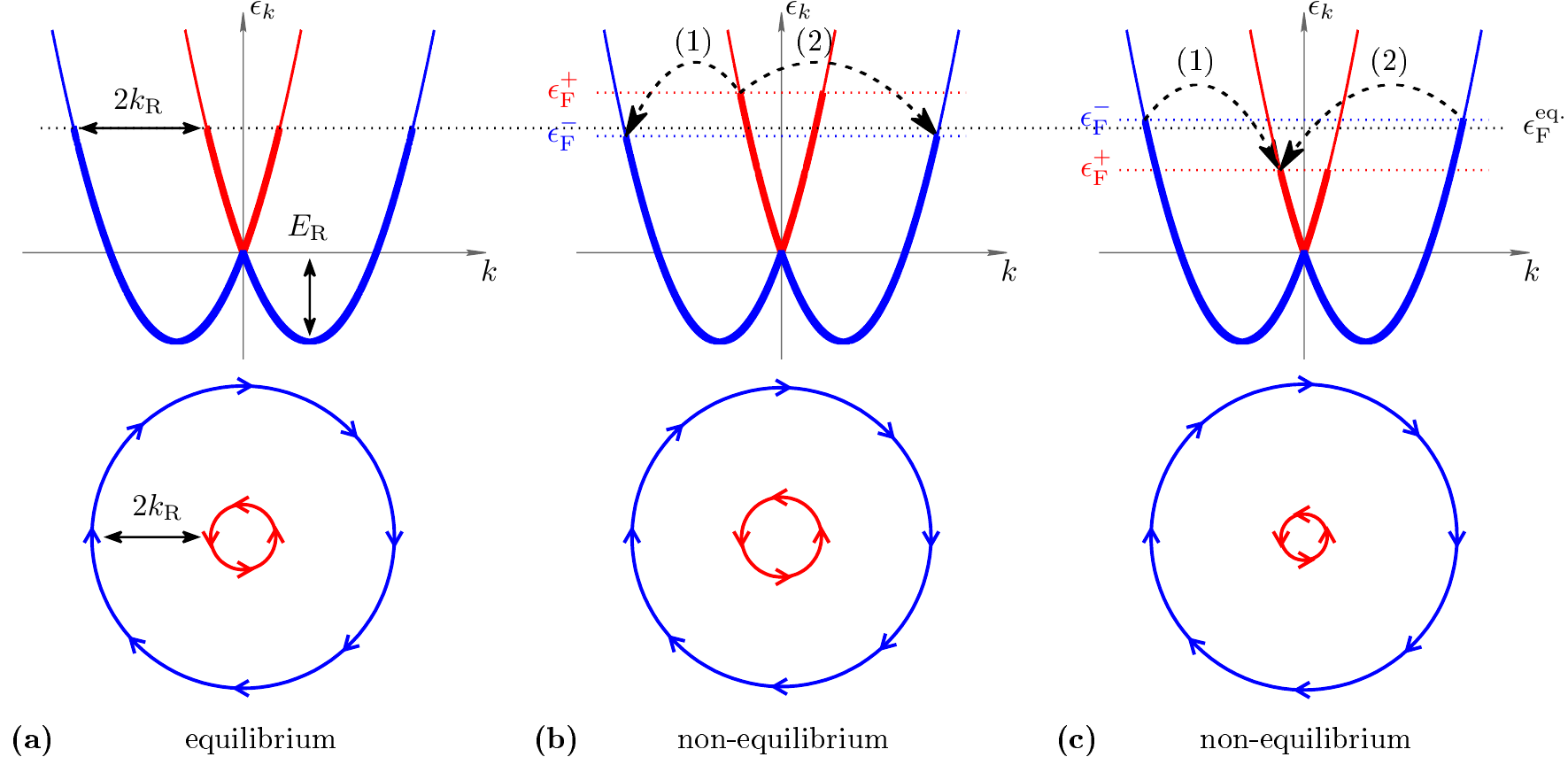}
  \caption{Equilibrium and non-equilibrium chirality distributions of charge carriers in a Rashba system. Top: energy dispersion with bold lines representing occupied states. Bottom: chiral spin structure at Fermi surface with arrows showing spin directions. \subfig{a} Equilibrium occupation of the Rashba dispersion. The spin-orbit coupling and lack of inversion symmetry lift spin degeneracy, resulting in two parabolic dispersions centred symmetrically away from $k = 0$. Consequently, Rashba systems have two concentric Fermi surfaces with opposite spin chirality that are separated in momentum space by $2\kR$. \subfig{b} Non-equilibrium occupation of the Rashba dispersion. The Fermi energy $\eF^+$ ($\eF^-$) of the upper (lower) Rashba band is higher (lower) than in equilibrium. In this example, the number of particles in the upper Rashba band chosen is $50\%$ higher than in equilibrium. These numbers are quite high, but were chosen for illustrative purposes. Equilibrium is restored via inter-band transitions, which must involve a spin flip or a reversal of momentum (labelled 1 and 2, respectively). \subfig{c} Vice versa of (b) with the number of particles in the upper band being $50\%$ less than in equilibrium.}
  \label{fig:fig01}
\end{figure*}

In this section we define the problem that we study, as well as the framework used to describe it. We introduce the Rashba model~\cite{Bihlmayer15}, which our analysis will be based on, and heuristically explain the nature of the long-lived chirality states that we study.

The breaking of spatial inversion symmetry lifts the spin degeneracy of a Bloch band, and leads to momentum-dependent spin mixing within the band. In the following, we will focus on a two-dimensional Rashba system, which is described by the Hamiltonian
\begin{equation}
  H = \frac{\hbar^2 \vk^2}{2m} - \aR \bm{\sigma}\cdot (\rso\times\vk) \label{eq:1}
\end{equation}
where $\hbar$ is Planck's constant, $\vk$ is the wavevector, $m$ is the effective mass of carriers, $\aR$ is the Rashba parameter, $\bm\sigma$ is the Pauli matrix vector, and $\rso$ is the direction of the spin-orbit coupling along which inversion symmetry is broken, which we assume to be orthogonal to the 2D system. The Hamiltonian in Eq. \eqref{eq:1} yields the energy dispersion $\ek^\pm = \nicefrac{\hbar^2}{2m} \, (k\pm\kR)^2 ~ - \ER$ with $\kR = \nicefrac{\aR m}{\hbar^2}$ and $\ER = \nicefrac{\hbar^2\kR^2}{2m}$, where the $+$ ($-$) superscript refers to the upper (lower) Rashba band, respectively. The dispersions and the resulting $k$-dependent spin alignments on the Fermi surface are shown in \figref[(a)]{fig01}. The free-electron parabolic dispersions of the two spin species are mutually shifted apart in momentum space by the Rashba momentum $\kR$. While the free electron system has two equal Fermi surfaces with degenerate spin and the same Fermi momentum, the Rashba system has a chiral spin texture, and two concentric Fermi surfaces at different Fermi momenta separated by $2\kR$.

We will investigate the relaxation of this system in a non-equilibrium state depicted in \figref[(b)]{fig01} and \figref[(c)]{fig01}. These show, respectively, charge carrier configurations in which the Fermi energy of the upper (lower) Rashba band, $\eF^+$ ($\eF^-$), is higher (lower) than in equilibrium, and vice versa. These states are clearly non-equilibrium states and must equilibrate on a finite period of time. In an ordinary electronic system, the relaxation of a non-equilibrium carrier population back to equilibrium can be expected to happen on very short timescales, typically on the order of picoseconds~\cite{Marder_2010}. In this work however, we provide evidence that the mechanisms relaxing this specific non-equilibrium distribution (i.e., chirality imbalance in strongly-coupled Rashba systems at low temperature) are suppressed, which results in much longer relaxation timescales. We note that such carrier populations could either be realised experimentally by injecting spin-polarised currents or by employing strong external magnetic fields (where this claim is further quantified in \secref{phonons}), however the focus of this contribution is not the experimental creation of these non-equilibrium states but rather a theoretical investigation of their relaxation.

Relaxation requires scattering events that induce transitions between the Rashba bands. The main process by which carriers can transition between bands is scattering off phonons as these can readily impart very large momentum and affect large-angle and backscattering effects. In addition to phonon scattering, momentum can also be imparted via inter-carrier scattering, and scattering from charged impurities. The present work discusses two key observations for such processes based on the energy-momentum conservation for phonons, and the nature of Coulomb scattering. In a system with strong Rashba coupling, the value of $\kR$ can be much larger than the thermal phonon momentum scale at low temperature $T$. Furthermore, scattering with charged impurities and inter-carrier Coulomb scattering is dominated by the transfer of small momenta, which for inter-band transitions is incompatible with the Rashba coupling because the spinor overlap vanishes for transitions with the smallest possible momentum transfer. These two points are discussed in greater detail in \secref{phonons} and \secref{coulomb}.

\section{Phonon Scattering}
\label{sec:phonons}

Charge carriers can exchange momentum with the solid either by emitting phonons, or by absorbing or scattering with thermally-excited phonons. These processes must obey energy-momentum conservation and therefore we have the following relations for the initial and final momenta, $\kk{}$ and $\kk{}' = \kk{} + \qq$, and corresponding energies, $\evk$ and $\evkp$:
\begin{equation*}
\arr{lccll}{
\mathrm{Emission} & \mathrm{:} & & \evk & \qquad = \qquad \evkp + \hbar \omega_{\mathbf{-q}} \punc{,}\\
\mathrm{Absorption} & \mathrm{:} & & \evk + \hbar \omega_{\mathbf{q}} & \qquad = \qquad \evkp \punc{,}\\
\mathrm{Scattering} & \mathrm{:} & & \evk + \hbar \omega_{\mathbf{p}} & \qquad = \qquad \evkp + \hbar \omega_{\mathbf{p} - \mathbf{q}} \punc{.}
}
\end{equation*}
Here $\hbar\omega_{\mathbf{q}}$ is the energy of a phonon with momentum $\qq$, and $\mathbf{p}$ is the momentum of a thermally-excited phonon. We assume the carrier initially to be in the band with higher Fermi energy (the inner one in \figref[(b)]{fig01} and the outer one in \figref[(c)]{fig01}), and define $\Delta\epsilon = \evk - \evkp$ to be the energy difference between initial and final states. We now combine energy conservation, Pauli exclusion, and the phonon dispersion to construct our argument for the suppression of these relaxation events in the case where $T$ is low such that,
\begin{equation}
    \kB T \ll \hbar \omega_{2\kR}. \label{eq:condT}
\end{equation}

\textit{Absorption}: for an inter-band transition, we require that $q \geq 2\kR$, however at low $T$ where \eqnref{condT} holds, such phonons are not thermally excited, thereby disallowing the absorption process.

\textit{Emission}: since for an allowed process $q \geq 2\kR$, we find $\hbar \omega_{2\kR} \leq \hbar \omega_{\mathbf{-q}} = \evk - \evkp = \Delta\epsilon$. Consequently, the Rashba momentum scale imposes an upper bound on the energy difference below which inter-band transitions are not allowed:
\begin{equation}
    \Delta\epsilon < \hbar \omega_{2\kR} \punc. \label{eq:cond}
\end{equation}
Therefore, the emission is prohibited as long as the energy difference $\Delta\epsilon$ between any occupied state in the Rashba band with higher Fermi energy and any empty state in the band with lower Fermi energy obeys above condition. This imposes a maximum detuning of the Fermi energies below which equilibration of the non-equilibrium state depicted in \secref{model} will be suppressed. We will discuss the relative strengths of the energy scales of $\Delta\epsilon$ and $\hbar\omega_{2\kR}$ in typical Rashba materials at the end of this section. Because this means that there are no unoccupied states to scatter into as the outcome of a phonon emission, one can think of this process as being Pauli-blocked by the occupied low-lying carrier states. Note that all thermally-activated excitations will be too small to disobey this condition as long as \eqnref{condT} holds.

Assuming an excess occupation in the Rashba bands of $n_\pm = n_\pm^{\mathrm{eq.}} (1\pm \delta)$, we find that for small $\delta$ condition \eqnref{cond} is satisfied when approximately
\begin{equation}
    \delta < \frac{1}{2} \frac{\hbar\omega_{2\kR}}{\eFeq+\ER} \punc. \label{eq:excess}
\end{equation}
The derivation of this and the exact expression are reported in \appref{excess}.

\textit{Scattering:} depending on the relative angle between $\mathbf{q}$ and $\mathbf{p}$, it is $\hbar \cph (q-2p) \leq \hbar \omega_{\mathbf{p}-\mathbf{q}} - \hbar \omega_{\mathbf{p}} = \Delta\epsilon \leq \hbar \cph q$, where we have assumed a linear acoustic phonon dispersion with speed of sound $\cph$ and that $p < q$. Consequently, the modified condition under which scattering is prohibited becomes
\begin{equation}
    \Delta\epsilon < \hbar \omega_{2(\kR-p)} \punc. \label{eq:cond2}
\end{equation}
When \eqnref{condT} holds, it is $p \ll \kR$, and therefore this condition is almost equivalent to the one for emission. Also note that the above only holds for the case where $\mathbf{p}$ and $\qq$ are antiparallel, hence the likelihood of such an event is already diminished in the first place.

It is easy to see how higher-order processes that are built up of several absorption and scattering events could eventually change the momentum of a carrier sufficiently to induce a transition into the other Rashba band while having no strong constraints on the energy difference between initial and final state. The contribution to equilibration can however be expected to be weak, not only because it is a higher-order process, but also because the phase space for such a process is small (the relative angles of phonon momenta have to be aligned in a particular way).

Furthermore, it is worth noting that our argument prohibits scattering events independently of whether they conserve or flip the spin of the charge carrier, as our discussion is purely based on energy-momentum conservations.

What is the degree to which these effects are present in physical systems? There are two relevant conditions to check, namely whether both $\kB T$ and $\Delta\epsilon$ are sufficiently less than $\hbar \omega_{2\kR}$. Simply speaking, the first condition ensures that there are no thermally-excited phonons to absorb for an inter-band transition, whereas the second ensures that the energy detuning is not so big that inter-band transitions can be induced by emission of phonons. For concreteness, we consider \ce{GeTe}, a system that is known to have giant Rashba coupling, in which $\kR = \SI{0.19}{\per\angstrom}$~\cite{Liebmann16}. Assuming a typical value of $\cph = \SI{3e+3}{\meter\per\second}$ and $T = \SI{1}{\kelvin}$, we find that $\kB T = \SI{0.09}{\milli\eV} \ll \hbar \omega_{2\kR} = 2 \hbar \cph \kR = \SI{7.5}{\milli\eV}$. To understand the implications of this on the relaxation timescales, note that the likelihood of such a relaxation event to take place is suppressed exponentially by the a Boltzmann factor of $\exp(\nicefrac{-2\hbar \cph \kR}{\kB T}) \approx \exp(-87) \approx 10^{-38}$, consequently resulting in effectively a complete suppression of any such phonon-induced relaxation events. Furthermore, condition \eqnref{excess} has to be satisfied, which yields $\delta < 1.6\%$ for GeTe with doping of $\eFeq = \nicefrac{\ER}{2}$ (i.e. close to but above the nodal crossing point, which is realistic for Ge-vacancy doping). Thus, the non-equilibrium state is protected from carrier-phonon relaxation as long as the population imbalance is not more than a few percent. Note that this number can change drastically depending on the Rashba coupling $\alpha$, the effective mass $m$, and the speed of sound $c$.

Before concluding this section, we use the derived expression for the Fermi level detuning in \appref{excess} and the upper limit $\delta$ found in this section to determine what magnetic field strength would be required to create carrier distributions with such chirality imbalances. Using
\begin{equation}
    g \muB B = \Delta \eF = 2\delta(\eF+\ER)
\end{equation}
and, using $g=2$, $\delta = 1.6\%$, $\eF = 0.5 \ER$, we find that $B \approx \SI{9.4}{\tesla}$. The absence of phonon relaxataion and the resulting long relaxation times will manifest for carrier imbalances created by magnetic fields below this value, whereas for larger fields the phonon relaxation may no longer be Pauli blockaded.

In summary, we have explained how in a Rashba system phonon-induced inter-band transitions of charge carriers are ineffective to relax a detuning of the Fermi level if the detuning is sufficiently small and the temperature is low, both compared to the energy scale $\hbar \omega_{2\kR}$, which is imposed by the Rashba momentum and the phonon dispersion. We have reasoned that this occurs because phonons are not available for absorption, the emission is Pauli blocked, and higher-order scattering will even for the best possible alignment of phonon momenta have only a comparatively small effect.

\section{Inter-carrier and Impurity-carrier Coulomb Scattering}
\label{sec:coulomb}

We continue by examining the role of scattering via the Coulomb interaction between charged carriers, and with charge localised impurities. In contrast to the phonon scattering case, here both the Rashba split $2\kR$ and the opposing helical spin structure at the Fermi surfaces (see \figref[(a)]{fig01}) play a role in the suppression of the transfer of small momenta (we mention that there are no restrictions on these processes \textit{within} a band, but this is irrelevant towards inducing inter-band transitions).

As indicated in \figref[(b)]{fig01} and \figref[(c)]{fig01}, inter-band transitions can broadly be split up into two types: (1) small-angle scattering events that change the momentum only marginally, but have to flip the spin, and (2) large-angle scattering that reverses the momentum and leaves the spin unchanged. We show that (1) is strongly suppressed because of the vanishing spinor overlap, which leaves processes of type (2) with large momentum transfer as the dominant mode of relaxation. Coulomb scattering however is dominated by the transfer of small momenta $q$ because the Fourier transform of the Coulomb potential is strong at $q\approx0$, and because (in the case of inter-carrier scattering) energy conservation is always satisfied when $q=0$, resulting in a logarithmically-divergent phase space. This incompatibility between the nature of the Coulomb interaction and the helical spin alignment of the Rashba energy dispersion is the reason why the Coulomb scattering is also suppressed. However, we also show in the following that this effect is much less pronounced than in the phonon case presented in the previous section.

The entirety of all scattering events includes cases in between (1) and (2), and the aim of the following analysis is to account for this. We derive expressions for the relaxation timescale of carrier-impurity and carrier-carrier scattering, which allows us to show how the above-mentioned arguments manifest quantitatively. Furthermore, we explicitly calculate the timescale for inter-carrier scattering for the case of GeTe. This section only reports the results, whereas detailed calculations can be found in \appref{app-time-constants}. We note that the calculations in this section exclude processes involving spinflips, whose role we discuss in \appref{spinflips}.

Our analysis will be based on Boltzmann transport theory, and we want to study the time dependence of the distribution function $\ff{\kk1}^\pm$, where the $+$ ($-$) superscript indicates the upper (lower) Rashba band index. We neglect external fields and temperature gradients, which allows us to reduce the Boltzmann equation to
\begin{equation}
    \pdb{f_{\kk1}^\pm}{t} = I_{\mathrm{ci}}[f_{\kk1}^\pm] + I_{\mathrm{cc}}[f_{\kk1}^\pm] \punc,
\end{equation}
where the indices ci and cc refer to the carrier-impurity and carrier-carrier contributions of the scattering integral, respectively. These are given by
\begin{align}
                        I_{\mathrm{ci}} = \sum_{\kk2}~~\big( & w_{(\kk2\mp) \rightarrow (\kk1\pm)}^{\mathrm{car-imp}} ~ f_{\kk2}^\mp [1 - f_{\kk1}^\pm] \nonumber \\[-8pt]
                      - & w_{(\kk1\pm) \rightarrow (\kk2\mp)}^{\mathrm{car-imp}} ~ f_{\kk1}^\pm [1 - f_{\kk2}^\mp] \big)
                        \label{eq:scattering_integral_ci}
\intertext{and}
                        I_{\mathrm{cc}} = \sum_{\mathclap{\kk2\kk3\kk4}}~~\big( & w_{(\kk3\kk4\mp)\to(\kk1\kk2\pm)}^{\mathrm{car-car}} ~ \ff{\kk3}^\mp \ff{\kk4}^\mp [1-\ff{\kk1}^\pm] [1-\ff{\kk2}^\pm] \nonumber\\[-8pt]
                      - & w_{(\kk1\kk2\pm)\to(\kk3\kk4\mp)}^{\mathrm{car-car}} ~ \ff{\kk1}^\pm \ff{\kk2}^\pm [1-\ff{\kk3}^\mp] [1-\ff{\kk4}^\mp] \big) \punc. \label{eq:scattering_integral_cc}
\end{align}

We have neglected all terms that conserve the Rashba band index of each particle or induce an exchange of particles between the bands, as these will not lead to a decay of the carrier imbalance between the two Rashba bands. We will make up for this by assuming that \textit{intra}-band scattering events are so quick that they will relax each individual band into \textit{local} thermal equilibrium on a timescale that is immediate compared to the \textit{inter}-band processes.

We calculate the probability amplitudes $w_{(\kk3\kk4\mp)\to(\kk1\kk2\pm)}$ using Fermi's Golden Rule,
\begin{align}
    w_{(\kk3\kk4\mp)\to(\kk1\kk2\pm)} = & \frac{2\pi}{\hbar} ~ |\bra{\Psi_{\mathrm{final}}} U \ket{\Psi_{\mathrm{init}}}|^2
    \label{eq:FermiGoldenRule}\\\nonumber
    & ~~~ \times ~ \delta(\ee1^\pm + \ee2^\pm - \ee3^\mp - \ee4^\mp) \punc,
\end{align}
for which we need to obtain the matrix element corresponding to the relevant transition. The matrix element will consist of two parts: 1) the Fourier transform of the Coulomb potential, which arises from its expectation value for the incoming and outgoing plane waves, and 2) the spinor overlap between initial and final states. We shall use the 2D Fourier transform of the screened 3D Coulomb potential for our calculations (where the static screening is obtained using the Random Phase Approximation)~\cite{Sarma15}, which reads
\begin{equation}
  U_{\pp} = U_p = \frac{2 \pi e_0^2}{p+\kS} \label{eq:coulombdef}
\end{equation}
with $e_0^2 = \nicefrac{e^2}{4 \pi \kappa \epsilon_0}$, where $\kappa$ is the effective background lattice dielectric constant and $\kS$ is the Thomas-Fermi screening momentum, and with $\pp = \kk{} - \kk{}'$, where $\kk{}$ and $\kk{}'$ are the incoming and outgoing plane waves. Furthermore, to determine the spinor overlap, we write each single-particle state as a superposition of the Pauli matrix $\sigma_{\mathrm{Z}}$ eigenstates~\cite{Smidman17} as
\begin{align}
    \ket{\kk{},+} &= \frac{1}{\sqrt2} \left( \ket{\kk{},\uparrow} - \I \expp{\I \theta_{\kk{}}} \ket{\kk{},\downarrow} \right), \\
    \ket{\kk{},-} &=  \frac{1}{\sqrt2} \left( - \I \expp{-\I \theta_{\kk{}}} \ket{\kk{},\uparrow} + \ket{\kk{},\downarrow} \right),
\end{align}
where $\theta_{\kk{}}$ is defined such that $\kk{} = (k_x, k_y)\tran = |\kk{}| ~ (\cos\theta_{\kk{}}, \, \sin\theta_{\kk{}})\tran$. This can then be used to evaluate the overlap between states from different bands, $\bra{\kk{}',+}U\ket{\kk{},-}$, and is equal to
\begin{equation}
 \expp{-\frac{\I}{2}\left(\theta_{\kk{}}+\theta_{\kk{}'}\right)} \sin\left( \nicefrac{ \theta_{\kk{}}-\theta_{\kk{}'} }{2}\right) \, \bra{\kk{}'} U \ket{\kk{}}. \label{eq:overlap}
\end{equation}

Following a well-known approach by Yafet~\cite{Yafet63}, which derives an expression for the spin relaxation time from phonon-assisted spinflip processes, we start from two Rashba bands, induce a small imbalance of the chemical potential, and perturb to first order in the detuning to find an expression for the relaxation time constants $\tau_{\mathrm{ci}}$ and $\tau_{\mathrm{cc}}$ for carrier-impurity and carrier-carrier scattering, respectively.

Our calculation yields the following expression for the relaxation time constant of the carrier-impurity processes.
\begin{equation}
    \frac{1}{\tau_{\mathrm{ci}}} = \frac{1}{\tau_{\mathrm{ci},0}} \frac{\pi}{4} \punc, \label{eq:previous_result_tau}
\end{equation}
where $\tau_{\mathrm{ci},0}^{-1} = {8\pi\,m\,n_{\rm i}\,e_0^4} \big/ {\hbar^3 (\kF^0)^2}$, $n_{\rm i}$ is the impurity density (per area), and $\kF^0$ is the average equilibrium Fermi momentum of the system. We compare this result to the case where we omit an essential feature of the Rashba system, namely the helical alignment of spin eigenstate at the Fermi surface, whose overlap is given by \eqnref{overlap}. This yields
\begin{equation}
    \frac{1}{\tau_{\mathrm{ci}}} = \frac{1}{\tau_{\mathrm{ci},0}} \frac{\kFO}{\kS} \punc,
\end{equation}
which is enhanced over the result in \eqnref{previous_result_tau} by a factor of $\frac{\kFO}{\kS}$. Since typically the Fermi momentum is much larger than the screening momentum scale, we see how the Rashba dispersion serves to enhance the screening of the Coulomb interaction. Using standard Lindhard theory, we estimate $\frac{\kFO}{\kS} \approx 23.7$ for GeTe. This result is worth noting, but equally not too relevant for practical applications, where the carrier-impurity scattering is mainly influenced by the impurity concentration $n_{\rm i}$ and the suppression by a factor of $\frac{\kFO}{\kS}$ will not be as big as in the case of electron-phonon scattering described above.

Furthermore, we derive the following expression for the carrier-carrier Coulomb scattering relaxation time constant,
\begin{equation}
\frac{1}{\tau_{\mathrm{cc}}} = \frac{1}{\tau_{\mathrm{cc},0}} \, \frac{(\kB T)^2}{(\mu-\ER)^2} \times \rho\left(\frac{\ER}{\mu}, \frac{\kB T}{\mu-\ER}\right) \punc,
\end{equation}
with
\begin{align}
\rho =~& \frac{\pi^2}{6} \Bigg(1-\frac{\ER}{\mu}\left(1-\log\left(\frac{\ER}{\mu}\right)\right) \nonumber \\
  & \, - \left(1-\frac{\ER}{\mu}\right) \, \log\left(\frac{\pi^2}{6}\frac{\kB T}{\mu-\ER}\right) \Bigg)
\end{align}
where $\tau_{cc,0}^{-1} = \frac{(2\pi e_0^2)^2}{2\pi\hbar} \, \frac{2m}{\hbar^2}$.

This result is of a form similar to the scattering lifetime of a quasi particle subject to inter-carrier Coulomb interaction reported by Zheng and Das Sarma~\cite{PhysRevB.53.9964}. The expression for $\rho$ contains a logarithmically-divergent part and a constant part, where the latter occurs due to the regularising effect of the opposing helical spin structure (which however only affects one part of the divergent phase space).

Using $T=\SI{100}{\milli\kelvin}$, as well as $\kappa = 10$, $\mu - \ER = \SI{0.1}{\electronvolt}$ and $\frac{\ER}{\mu} \approx 0.5$ (which we assume for 32~nm thick $\alpha$-GeTe~\cite{THE67,Liebmann16, Narayan16}), we find
\begin{equation}
    \tau_{\mathrm{cc}} \approx \SI{1}{\micro\second} \punc.
\end{equation}

This result is impressive in that the lifetime is significantly enhanced over the picosecond lifetime that is commonly observed in electronic systems~\cite{Marder_2010}. Normally relaxation timescales on the order of microseconds are only observed for carrier-phonon processes at low temperature \cite{PhysRevLett.99.145503}. However, we also note that the suppression is not of the exponential form found for the phonon case, and hence not as dramatic. Furthermore, we note that the chiral alignment of spin eigenstates on the Fermi surface only serves to suppress one of the two logarithmic divergences at $q=2\kR$ of the form $\log(\nicefrac{\mu-\ER}{\kB T})$ and does not affect the same divergence occuring at $q = \kF^+ + \kF^-$. It would be interesting to see if more complex band structures involving more intricate spin alignments could result in even stronger suppression and even longer relaxation timescales.

We note that, while our calculation is sufficiently accurate to approximate the order of magnitude of the relaxation time constant, an estimate of the time constants for real materials necessitates the consideration of exact band structures and possibly dynamic screening and exchange interaction effects.

\section{Discussion and Conclusions}
\label{sec:conclusions}
In conclusion, we have studied the relaxation of chirality imbalances in Rashba systems, and the relaxation processes and timescales associated with those states. We have shown that phonon-mediated inter-band transitions in Rashba systems are effectively absent when $T$ is low and the occupation number detuning $\delta$ is small. This happens due to an energy-momentum mismatch between the electronic and phonon dispersion, and consequently phonons do not contribute to relaxation of carrier chirality. We identify the inter-carrier scattering mediated by the Coulomb interaction as the resulting dominant relaxation mechanism (in a pure sample) and further analyse it. We find that this is also weakened due to the chiral spin structure at the Fermi level, and consequently the relaxation time arising from this is much longer than what is commonly expected from inter-carrier interaction. We estimate the relaxation timescales for the inter-carrier scattering for a typical strongly Rashba-coupled System (\ce{GeTe} in our example) at low temperatures of $T\approx\SI{100}{\milli\kelvin}$ to be on the order of $\tau\approx\SI{1}{\micro\second}$.

We note that these results mainly rely on the large momentum split that strong Rashba coupling induces between the Fermi surfaces of the bands of the two carrier species, as well as on the fact that carriers from different Rashba bands with lowest momentum separation have orthogonal spin states (which is always true for systems with a spherical Fermi surface). As such, it is possible to generalise the results from this study to systems with different energy dispersion, as long as those two main ingredients are retained.

More generally, a remarkable finding of our study is that due to the absence of phonons the dominant mode of inter-band relaxation is carrier-carrier Coulomb scattering, whereas commonly this is mediated by phonons.

The non-equilibrium chirality populations we study could be realised in experiments in a number of ways, such as through the application of magnetic fields, as well as by injecting spin-polarised currents that dominantly occupy one of the two Rashba bands, although the analysis of this is beyond the scope of this manuscript and requires further work.

\begin{acknowledgments}
PCV acknowledges useful discussions with Victor Jouffrey, Bartholomew Andrews, Chris Hooley, James Dann, Gareth Conduit, Giulio Schober, Ivona Bravic, Jan Behrends, and Be\~nat Mencia. VN acknowledges funding from the Engineering and Physical Sciences Research Council (EPSRC), UK.
\end{acknowledgments}

\appendix

\section{Condition on Excess Occupation}
\label{sec:excess}
This appendix section aims to briefly derive an expression for the upper bound of the excess occupation $\delta$ from the Rashba band dispersion $\e{k}$ and the value of the equilibrium Fermi energy $\eFeq$.

First, we compute the value of the Fermi momentum for any given Fermi energy $\eF$. We define the free energy dispersion $\ekf = \nicefrac{\hbar^2k^2}{2m}$, and by setting $\eFpm = \e{\kF}^\pm$ and solving for $\ekfFpm$ we find
\begin{equation}
    \ekfFpm = \eFpm + 2\ER \pm 2\sqrt{\ER^2+\ER\eFpm} \punc.
\end{equation}
The total number of carriers $n_\pm$ is given by
\begin{align}
n_\pm &= \int\!\frac{\ddd{k}{2}}{(2\pi)^2} \, \Theta(\eF - \ek^\pm)\,, \\
\intertext{where $\Theta(x)$ is the Heaviside function. By substituting $\epsilon = \ekf$,}
n_\pm &= \frac{m}{2\pi\hbar^2} \int_0^\infty\!\!\!\dd{\epsilon}~
         \Theta(\ekfFpm - \epsilon) = \frac{m}{2 \pi\hbar^2} \ekfFpm \punc.
\end{align}
We define the reduced carrier density $\nu^\pm = \frac{2\pi\hbar^2}{m} n^\pm$, which has the dimension of energy, such that
\begin{equation}
    \nu_\pm = \eFpm + 2\ER \pm 2\sqrt{\ER^2+\ER\eFpm} \punc.
\end{equation}
This equation can be inverted to yield the Fermi energies $\eFpm$ from the carrier density $\nu_\pm$,
\begin{equation}
    \eFpm = \nu_\pm \pm 2\sqrt{\ER \nu_\pm} \punc.
\end{equation}
Therefore, given occupations $\nu_\pm = \nu_\pm^{\mathrm{eq.}} (1\pm \delta)$, we can calculate the Fermi energy difference to be
\begin{align}
    \Delta\eF  &= \eF^+ - \eF^- \\
               &= \nu_+^{\mathrm{eq.}} (1+\delta) - \nu_-^{\mathrm{eq.}} (1-\delta) \\
               &~~~~+2\sqrt{\ER}\left(\sqrt{\nu_+^{\mathrm{eq.}}(1+\delta)}+\sqrt{\nu_-^{\mathrm{eq.}}(1-\delta)}\right) \nonumber \\
               &= 2\delta (\eFeq+2\ER) - 4\sqrt{\ER^2+\eFeq\ER} + 2\sqrt{\ER} \\
               &~~~~\times\bigg[\sqrt{1+\delta}\sqrt{\eFeq+2\ER-2\sqrt{\ER^2+\eFeq\ER}} \nonumber \\
               &~~~~~\,+\sqrt{1-\delta}\sqrt{\eF+2\ER+2\sqrt{\ER^2+\eF\ER}}\bigg], \nonumber \\
\intertext{and by expanding $\sqrt{1+\delta}+\sqrt{1-\delta}$ as a series around $\delta=0$, we find}
               &= 2\delta (\eF+\ER) + 4\sqrt{\ER} \\
               &~~~~\times\bigg[\sqrt{\ER+\eFeq} \sum_{k=1}^\infty \left(\arr{c}{\nicefrac{1}{2}\\2k} \right) \delta^{2k}\nonumber\\
               &~~~~~~~~~-\, \sqrt{\ER} ~~~~\, \sum_{k=1}^\infty \left(\arr{c}{\nicefrac{1}{2}\\2k+1} \right) \delta^{2k+1}\bigg]. \nonumber
\end{align}
When $\delta$ is small, the series expansion can be truncated after the first element, which yields the result reported in \secref{phonons}.

\section{Relaxation Time Constants}
\label{sec:app-time-constants}

\subsection{General Considerations}

In this appendix, we provide a detailed explanation of our calculations of the Boltzmann transport scattering integrals that are discussed in \secref{coulomb} of the main text. We start by providing some general calculations that hold for carrier-impurity and carrier-carrier scattering equally and continue by calculating expressions for the relaxation time constants explicitly for the two cases.

\subsubsection{Boltzmann Equation and Scattering Integrals}
We consider the Boltzmann equation,
\begin{equation}
    \pdb{f_{\kk1}^\pm}{t} = I_{\mathrm{ci}}[f_{\kk1}^\pm] + I_{\mathrm{cc}}[f_{\kk1}^\pm] \punc,
\end{equation}
where the indices ci and cc refer to the carrier-impurity and carrier-carrier contributions of the scattering integral, respectively. These are given by
\begin{align}
                        I_{\mathrm{ci}} = \sum_{\kk2}~~\big( & w_{(\kk2\mp) \rightarrow (\kk1\pm)}^{\mathrm{car-imp}} ~ f_{\kk2}^\mp [1 - f_{\kk1}^\pm] \nonumber \\[-8pt]
                      - & w_{(\kk1\pm) \rightarrow (\kk2\mp)}^{\mathrm{car-imp}} ~ f_{\kk1}^\pm [1 - f_{\kk2}^\mp] \big)
                        \label{eq:sup:scattering_integral_ci}
\intertext{and}
                        I_{\mathrm{cc}} = \sum_{\mathclap{\kk2\kk3\kk4}}~~\big( & w_{(\kk3\kk4\mp)\to(\kk1\kk2\pm)}^{\mathrm{car-car}} ~ \ff{\kk3}^\mp \ff{\kk4}^\mp [1-\ff{\kk1}^\pm] [1-\ff{\kk2}^\pm] \nonumber\\[-8pt]
                      - & w_{(\kk1\kk2\pm)\to(\kk3\kk4\mp)}^{\mathrm{car-car}} ~ \ff{\kk1}^\pm \ff{\kk2}^\pm [1-\ff{\kk3}^\mp] [1-\ff{\kk4}^\mp] \big) \punc. \label{eq:sup:scattering_integral_cc}
\end{align}
We have neglected all terms that conserve the Rashba band index of each particle or induce an exchange of particles between the bands (as the latter will not lead to a decay of the carrier imbalance between the two Rashba bands). We will make up for this by assuming that \textit{intra}-band scattering events are so quick that they will relax each individual band into \textit{local} thermal equilibrium on a timescale that is immediate compared to the \textit{inter}-band processes we study. We note that there exist additional inter-carrier scattering processes that conserve the Rashba band index of one carrier but not of the other. These however are (due to considerations based on momentum conservation) absent when $\kR > \kF^+$, and remain weak as long as $\kR \approx \kF^+$, which we assume to be the case and which is the case for Rashba materials where the Rashba energy scale is comparable to the Fermi energy.

\subsubsection{Defining the Chirality Density}
We define the total particle density, $n = n_+ + n_-$, and the total chirality density, $C = n_- - n_+$ (which is defined to be a positive number, as there are more particles in the lower Rashba band). Using the distribution function $f^\pm_{\kk{}}$, we can write the densities as $n_\pm = \frac{1}{\vol} \sum_{\kk1} \ff{\kk1}^\pm$, and therefore we can express $C$ and its derivative with respect to time $t$ as,
\begin{align}
    C &= \frac{1}{\vol} \sum_{\kk1} \left( \ff{\kk1}^- - \ff{\kk1}^+ \right) \punc, \label{eq:sup:defC}\\
    \frac{\dd{C}}{\dd{t}} &= \frac{1}{\vol} \sum_{\kk1} \left( \pdb{\ff{\kk1}^-}{t} - \pdb{\ff{\kk1}^+}{t} \right) \punc. \label{eq:sup:defdCbydt}
\end{align}
where $\vol$ is the area of the system.

We insert the carrier-impurity contribution of the scattering integral in \eqnref{sup:scattering_integral_ci} and the carrier-carrier contribution in \eqnref{sup:scattering_integral_cc} into \eqnref{sup:defdCbydt}. By, in the second term, renaming $\kk1 \mapsto \kk2$ and $\kk2 \mapsto \kk1$ for the carrier-impurity case and $\kk1,\kk2 \mapsto \kk3,\kk4$ and $\kk3,\kk4 \mapsto \kk1,\kk2$ in the carrier-carrier case, and then making use of the reversibility of the microscopic processes,
\begin{align}
    w_{(\kk2+) \rightarrow (\kk1-)}^{\mathrm{car-imp}} &= w_{(\kk1-) \rightarrow (\kk2+)}^{\mathrm{car-imp}} \\
    w_{(\kk3\kk4+)\to(\kk1\kk2-)}^{\mathrm{car-car}} &= w_{(\kk1\kk2-)\to(\kk3\kk4+)}^{\mathrm{car-car}}
\end{align}
we find
\begin{align}
    \frac{\dd{C}}{\dd{t}} \bigg|_{\mathrlap{\mathrm{ci}}}~ = \frac{2}{\vol} ~~\sum_{\mathclap{\kk1\kk2}}~ & w_{(\kk1-)\to(\kk2+)}^{\mathrm{car-imp}} (\ff{\kk2}^+ - \ff{\kk1}^-) \punc. \label{eq:sup:dCbydtci} \\
    \frac{\dd{C}}{\dd{t}} \bigg|_{\mathrlap{\mathrm{cc}}}~ = \frac{2}{\vol} ~~\sum_{\mathclap{\kk1\kk2\kk3\kk4}}~ & w_{(\kk1\kk2-)\to(\kk3\kk4+)}^{\mathrm{car-car}} \label{eq:sup:dCbydtcc} \\[-8pt]
    & \qquad \quad~ \Big( \ff{\kk3}^+ \ff{\kk4}^+ [1-\ff{\kk1}^-] [1-\ff{\kk2}^-] \nonumber\\[-5pt]
    & \qquad - \hphantom{\Big(}\ff{\kk1}^- \ff{\kk2}^- [1-\ff{\kk3}^+] [1-\ff{\kk4}^+] \Big) \punc. \nonumber
\end{align}

\subsubsection{Inducing a Fermi-level Detuning}
We use the Rashba energy dispersion $\ek^\pm = \frac{\hbar^2(k\pm\kR)^2}{2m} - \ER$ (where $\ER=\nicefrac{\hbar^2\kR^2}{2m}$ is the Rashba energy) as shown in \figref{fig01}, although we will shift all energies by $\ER$ so that we can neglect the offset and assume the entire energy dispersion to take positive values.

We assume the equilibrium Fermi momentum $\kF^{\mathrm{eq.}\pm} = \kF^0 \mp \kR$ and the equilibrium chemical potential $\mueq = \nicefrac{\hbar^2 (\kF^0)^2}{2m}$. A chirality non-equilibrium distribution that conserves the total density $n = n_+ + n_-$ is induced by letting $\left(\kF^\pm\right)^2 = \left(\kF^{\mathrm{eq.}\pm}\right)^2 \mp \left(\delta k \vphantom{\kF^\pm}\right)^2$. By defining $\dmu = \nicefrac{\hbar^2(\delta k)^2}{2m}$, we can express the non-equilibrium chemical potentials as
\begin{equation}
    \label{eq:sup:mu}
    \mu^\pm = \mueq \mp \frac{1}{1\mp\sqrt{\frac{\ER}{\mueq}}} \dmu + O(\dmu^2) \punc{.}
\end{equation}

We replace the distribution functions $\ff{\kk{}}^\pm$ in Eqs.~\eqref{eq:sup:defC}, \eqref{eq:sup:dCbydtci}, and \eqref{eq:sup:dCbydtcc} by Fermi distributions $\fermi{\mu^\pm}{\ee{}^\pm}$ with chemical potential $\mu^\pm$, which can, by using \eqnref{sup:mu}, be expanded to first order in $\dmu$ as
\begin{align}
    \fermi{\mu^\pm}{\epsilon} &= \f{\epsilon} \mp \chi^\pm \dmu \pdb{\f{\epsilon}}{\mu}\punc{,} \nonumber\\
    [1-\fermi{\mu^\pm}{\epsilon}] &= [1-\f{\epsilon}] \mp \chi^\pm \dmu \pdb{[1-\f{\epsilon}]}{\mu} \punc{,} \nonumber
\end{align}
where we have defined $\chi^\pm = (1\mp\sqrt{\nicefrac{\ER}{\mueq}})^{-1}$, and $f$ is the equilibrium Fermi distribution. By using $\pdb{f}{\mu} = -\pdb{f}{\epsilon} = \beta \, \f{\e{}} \, [1-\f{\e{}}]$ and $\pdb{[1-\f{\e{}}]}{\mu} = -\beta \, \f{\e{}} \, [1-\f{\e{}}]$, we can rewrite $C$ and $\nicefrac{\dd{C}}{\dd{t}}$ as
\begin{align}
  C - \Ceq &= + ~\, \frac{\dmu\,\beta}{\vol} ~\, ~\sum_{\kk1}~ \Phi^0(\ee1^-, \ee1^+) \\
  \dCbydt \bigg|_{\mathrlap{\mathrm{ci}}}~ &= - \frac{2 \, \dmu \, \beta}{\vol} ~ ~\sum_{\mathclap{\kk1\kk2}}~ \Phi^0(\ee1^-, \ee2^+) \\[-8pt]
           & \qquad\qquad\qquad\qquad~~ \times w_{(\kk1-)\to(\kk2+)}^{\mathrm{car-imp}} \nonumber \\
  \dCbydt \bigg|_{\mathrlap{\mathrm{cc}}}~ &= - \frac{2 \, \dmu \, \beta}{\vol} ~ ~\sum_{\mathclap{\kk1\kk2\kk3\kk4}}~ \Phi^1(\ee1^-, \ee2^-, \ee3^+, \ee4^+) \\[-8pt]
           & \qquad\qquad\qquad\qquad~~ \times w_{(\kk1\kk2-)\to(\kk3\kk4+)}^{\mathrm{car-car}}
\punc.\nonumber
\end{align}
The zeroth order term in $\dmu$ for $\dCbydt$ vanishes as the equilibrium carrier configuration does not induce any changes of $C$. Furthermore, we have introduced $\Phi^0(\e1, \e2)$, and $\Phi^1(\e1,\e2,\e3,\e4)$, which account for all distribution functions,
\begin{align}
\Phi^0 &= \chi^- \f{\e1}[1-\f{\e1}]
+ \chi^+ \f{\e2}[1-\f{\e2}] \\
\Phi^1 &= ~~\f{\e3} \f{\e4} [1-\f{\e1}] [1-\f{\e2}] \\
       &\qquad\qquad\times \Big( \chi^+ \left([1-\f{\e3}] + [1-\f{\e4}]\right) \nonumber\\
       &\qquad\qquad~~+ \chi^- \left(\f{\e1} + \f{\e2}\right)\Big) \nonumber \\
       &+ ~~\f{\e1} \f{\e2} [1-\f{\e3}] [1-\f{\e4}] \nonumber\\
       &\qquad\qquad\times \Big( \chi^+ \left( \f{\e3} + \f{\e4} \right) \nonumber\\
       &\qquad\qquad~~+ \chi^- \left( [1-\f{\e1}] + [1-\f{\e2}] \right) \Big) \nonumber
\end{align}
The factors of $\chi\pm$, which account for the difference in density of states of the two Rashba bands at their respective Fermi energies, can be approximated as $1$, and we will therefore neglect them in the following.

Using the expressions for $C$ and $\frac{\dd{C}}{\dd{t}}$, we want to find the relaxation time constant $\tau$ in the following relaxation time approximation:
\begin{equation}
    \frac{\dd{C}}{\dd{t}} = - \frac{C-\Ceq}{\tau} \punc{.}
\end{equation}
This yields
\begin{align}
    \label{eq:sup:exprtauci}
    \frac{1}{\tau_{\mathrm{ci}}} = & \,\frac{2}{\vol} ~~\sum_{\mathclap{\kk1\kk2}}~ \Phi^0(\ee1^-, \ee2^+) \times w_{(\kk1-)\to(\kk2+)}^{\mathrm{car-imp}} \nonumber \\
     & \qquad\qquad \Bigg/ \frac{1}{\vol} \sum_{\kk1} \Phi^0(\ee1^-, \ee1^+) \punc, \\
    \label{eq:sup:exprtaucc}
    \frac{1}{\tau_{\mathrm{cc}}} = & \,\frac{2}{\vol} ~~\sum_{\mathclap{\kk1\kk2\kk3\kk4}}~ \Phi^1(\ee1^-, \ee2^-, \ee3^+, \ee4^+) \times w_{(\kk1\kk2-)\to(\kk3\kk4+)}^{\mathrm{car-car}} \nonumber \\
     & \qquad\qquad \Bigg/ \frac{1}{\vol} \sum_{\kk1} \Phi^0(\ee1^-, \ee1^+) \punc.
\end{align}

\subsubsection{Fermi's Golden Rule}
The probability amplitudes in Eqns.~\eqref{eq:sup:exprtauci} and \eqref{eq:sup:exprtaucc} are given by Fermi's Golden Rule,
\begin{align}
    w_{(\kk3\kk4\mp)\to(\kk1\kk2\pm)} = & \frac{2\pi}{\hbar} ~ |\bra{\Psi_{\mathrm{final}}} U \ket{\Psi_{\mathrm{init}}}|^2
    \label{eq:sup:FermiGoldenRule}\\\nonumber
    & ~~~ \times ~ \delta(\ee1^\pm + \ee2^\pm - \ee3^\mp - \ee4^\mp) \punc,
\end{align}
where the subscripts init and fin refer to the initial and final states of the transition. We now have to find the matrix element for the respective transitions.

Initial and final states are taking the forms
\begin{align}
    & \ket{\Psi_{\mathrm{init}}^\mathrm{ci}} & \!\!\! & = & \!\!\! & \ket{\kk1, -} \punc,  & &
    \ket{\Psi_{\mathrlap{\mathrm{fin}}\hphantom{\mathrm{init}}}^\mathrm{ci}} & \!\!\! & = & \!\!\! & \ket{\kk2, +} \punc, \nonumber \\[5pt]
    & \ket{\Psi_{\mathrm{init}}^\mathrm{cc}} & \!\!\! & = & \!\!\! & \ket{\kk1,-} \ket{\kk2,-} \punc,
    & &
    \ket{\Psi_{\mathrlap{\mathrm{fin}}\hphantom{\mathrm{init}}}^\mathrm{cc}} & \!\!\! & = & \!\!\! & \ket{\kk3,+} \ket{\kk4,+} \punc.
\end{align}
We note that a full treatment of this problem would envolve an antisymmetrisation of the two-particle wave-function, which would result in exchange-interaction terms in the matrix element. This however makes the integration that follows further below intractable. We also do not expect this to have a big impact onto the end result as exchange interaction only matters when a process and its exchange process are of similar strength, which is not true for the parts of the phase space that contribute dominantly to the scattering.

Each single-particle state can be written as a superposition of the Pauli matrix $\sigma_{\mathrm{Z}}$ eigenstates~\cite{Smidman17} as follows:
\begin{align}
    \ket{\kk{},+} &= \frac{1}{\sqrt2} \left( \ket{\kk{},\uparrow} - \I \expp{\I \theta_{\kk{}}} \ket{\kk{},\downarrow} \right) \punc, \\
    \ket{\kk{},-} &=  \frac{1}{\sqrt2} \left( - \I \expp{-\I \theta_{\kk{}}} \ket{\kk{},\uparrow} + \ket{\kk{},\downarrow} \right) \punc.
\end{align}
Because the component of $U$ in spin space is the identity, its matrix element, $\bra{\kk{}',+} U \ket{\kk{},-}$, can be written as
\begin{align}
     & \, \frac{1}{2} \Big(\!\bra{\kk{}',\uparrow} \!+\! \I \expp{-\I\theta_{\kk{}'}} \bra{\kk{}',\downarrow} \Big) U \Big( \!-\! \I \expp{-\I \theta_{\kk{}}} \ket{\kk{},\uparrow} \!+\! \ket{\kk{},\downarrow} \Big) \\
    =& \, \frac{1}{2} \left( - \I \expp{-\I \theta_{\kk{}'}} + \I \expp{-\I \theta_{\kk{}}}\right) \, \bra{\kk{}'}U\ket{\kk{}} \\
    =& \, \frac{1}{2\I} \expp{-\frac{\I}{2}\left(\theta_{\kk{}}+\theta_{\kk{}'}\right)} \left( \expp{\frac{\I}{2}\left(\theta_{\kk{}}-\theta_{\kk{}'}\right)} - \expp{-\frac{\I}{2}\left(\theta_{\kk{}}-\theta_{\kk{}'}\right)}\right) \, \bra{\kk{}'}U\ket{\kk{}} \\
    =& \, \expp{-\frac{\I}{2}\left(\theta_{\kk{}}+\theta_{\kk{}'}\right)} \sin\left( \nicefrac{ \theta_{\kk{}}-\theta_{\kk{}'} }{2}\right) \, \bra{\kk{}'}U\ket{\kk{}} \punc{.} \label{eq:sup:spinoroverlap}
\end{align}

We set $\pp = \kk{} - \kk{}'$, and hence $\bra{\kk{}'}U\ket{\kk{}} = {U_{\pp}}/{\vol}$, where $\vol$ is the area of the system and $U_{\pp}$ is the Fourier transform of the Coulomb potential~\cite{Sarma15},
\begin{equation}
  U_{\pp} = U_p = \frac{2 \pi e_0^2}{p+\kS} \label{eq:sup:coulombdef}
\end{equation}
with $e_0^2 = \nicefrac{e^2}{4 \pi \kappa \epsilon_0}$, where $\kappa$ is the effective background lattice dielectric constant and $\kS$ is the Thomas-Fermi screening momentum.

\subsection{Impurity Scattering}
Next, we evaluate the time constant for carrier-impurity scattering, which is easy to compute as carrier-impurity scattering is a single-particle process that conserves energy.

Following the general considerations from the previous section, we find that $w_{(\kk1-)\to(\kk2+)}^{\mathrm{car-imp}}$ is equal to
\begin{equation}
    \frac{2\pi}{\hbar \vol^2} \, \sin^2\left( \nicefrac{ \theta_{\kk1}-\theta_{\kk2} }{2}\right) \, U_{|\kk1-\kk2|}^2 \, \delta(\ee1^- - \ee2^+) \punc,
\end{equation}
which we will insert into \eqnref{sup:exprtauci} to obtain an expression for the time constant. We convert the sums to integrals to find
\begin{align}
    \label{eq:sup:exprtauci_sumstoints}
    & \frac{1}{\tau_{\mathrm{ci}}} = \,\frac{4\pi}{\hbar\vol} \iint \frac{\dd{\kk1}\dd{\kk2}}{(2\pi)^4}~ \Phi^0(\ee1^-, \ee2^+) \, \delta(\ee1^- - \ee2^+) ~ \times \nonumber \\
     & ~~~~U_{|\kk1-\kk2|}^2 \, \sin^2\left( \nicefrac{ \theta_{\kk1}-\theta_{\kk2} }{2}\right) ~\mathclap{\Bigg/}~~\mathclap{\int}\, \frac{\dd{\kk1}}{(2\pi)^2}\,\Phi^0(\ee1^-, \ee1^+) \punc.
\end{align}
We use the variable transformation $k_i \dd{k_i} = \frac{2m}{\hbar^2} \, \dd{\e{i}^\pm}\,\frac{1}{2}\,(1\pm\sqrt{\nicefrac{\ER}{\e{i}}})$ (for $i=1,2$). Similar to our previous approximation of assuming $\chi^\pm\approx1$, we drop the correction in the brackets, as this only accounts for a small correction stemming from the different densities of states of the two Rashba bands. Furthermore, we take the limit $T\to0$, in which case $\Phi^0(\e1,\e2) \to \beta^{-1} (\delta(\e1-\mu)+\delta(\e2-\mu))$, to carry out the integrations over $\e1$ and $\e2$.

This yields for the denominator
\begin{align}
 & \int \frac{\dd{\kk1}}{(2\pi)^2}\,\Phi^0(\ee1^-, \ee1^+) = \frac{1}{2\pi} \frac{2m}{\hbar^2} \, \beta^{-1} \punc.
\end{align}
Using this and carrying out the integration over $\e1$ and $\e2$ in the numerator, the result for timescale can be expressed as
\begin{align}
    & \frac{1}{\tau_{\mathrm{ci}}} = \,\frac{1}{2\pi\hbar\vol} \frac{2m}{\hbar^2} \int \dd{\theta_{\kk1,\kk2}} \, U_{|\kk1-\kk2|}^2 \, \sin^2\left( \nicefrac{ \theta_{\kk1}-\theta_{\kk2} }{2}\right) \punc,
\end{align}
where $\theta_{\kk1,\kk2}$ is the angle between $\kk1$ and $\kk2$. By writing $|\kk{\nicefrac{1}{2}}| = \kF^{\mathrm{eq.}\pm} = \kF^0 \mp \kR$, we can derive that $|\kk1-\kk2|^2 = 4 (\kFO)^2 \sin^2(\nicefrac{\theta}{2}) + 4 \kR^2 \cos^2(\nicefrac{\theta}{2}) \approx 4 (\kFO)^2 \sin^2(\nicefrac{\theta}{2}) + 4 \kR^2$. Using this, we can write
\begin{align}
\frac{1}{\tau_{\mathrm{ci}}} =
 ~& \frac{1}{\tau_{\mathrm{ci},0}} \int_{\mathrlap{0}}^{\mathrlap{\pi}} \! \dd{\theta} \frac{\sin^2\left( \nicefrac{ \theta}{2}\right)}{\bigg(2\sqrt{\sin^2(\nicefrac{\theta}{2}) \!+\! \left(\frac{\kR}{\kFO}\right)^2}+\frac{\kS}{\kFO}\bigg)^2} , \\
\intertext{where we have also multiplied the result with the number of impurity sites in the system $N_{\rm i}$, and introduced $n_{\rm i}=\nicefrac{N_{\rm i}}{\vol}$ and $\tau_{\mathrm{ci},0}^{-1} = {8\pi\,m\,n_{\rm i}\,e_0^4} \big/ {\hbar^3 (\kF^0)^2}$. In the last expression, we can observe that $\kR$ appears in a similar way to the screening momentum $\kS$. Because in strongly Rashba-coupled systems $\kR$ is several orders of magnitude larger than $\kS$, this therefore serves to enhance the screening of the Coulomb potential. More importantly, in the case without Rashba coupling, it is $\kR=0$ and the spinor overlap matrix element vanishes, and therefore}
\frac{1}{\tau_{\mathrm{ci}}} =
 ~& \frac{1}{\tau_{\mathrm{ci},0}} \int_{\mathrlap{0}}^{\mathrlap{\pi}} \dd{\theta} \frac{1}{\left(2 \sin(\nicefrac{\theta}{2})+\frac{\kS}{\kFO}\right)^2} \approx \frac{1}{\tau_{\mathrm{ci},0}} \frac{\kFO}{\kS} \punc,
\intertext{whereas if instead we retain $\kR$ and include the helical spin structure and use the fact that $\kS \ll \kR$, we instead find}
\frac{1}{\tau_{\mathrm{ci}}} =
 ~& \frac{1}{\tau_{\mathrm{ci},0}} \int_{\mathrlap{0}}^{\mathrlap{\pi}} \dd{\theta} \frac{1}{4} \frac{\sin^2(\nicefrac{\theta}{2})}{\sin^2(\nicefrac{\theta}{2}) + \left(\frac{\kR}{\kFO}\right)^2} \approx \frac{1}{\tau_{\mathrm{ci},0}} \frac{\pi}{4}.
\end{align}
This results in an overall suppression factor of $\approx \nicefrac\kS\kFO$. We use Lindhard theory~\cite{Marder_2010}, which gives $\kS = \frac{e^2 m}{\hbar^2}$, and using the effective mass $m \approx 0.02 m_{\rm e}$ ($m_e$ being the electron mass), we find $\nicefrac\kS\kFO \approx 23.7$, as reported in \secref{coulomb}. While this suppression is not as dramatic as in the phonon-scattering and inter-carrier cases, we note that to obtain the effective rate for carrier-impurity scattering, this suppression factor must be multiplied by a density of impurities. Therefore, in clean, undoped samples, in which the concentration of unintentional dopants is negligibly small, we expect this not to be of importance.

\subsection{Inter-carrier Scattering}
We now continue with examining the carrier-carrier contribution to relaxation, which is harder and requires significantly more work.

\subsubsection{Inserting Fermi's Golden Rule and Converting Sums to Integrals}
Using the result in Eqns.~\eqref{eq:sup:spinoroverlap} and \eqref{eq:sup:coulombdef}, the matrix element squared, $|\bra{\Psi_{\mathrm{final}}} U \ket{\Psi_{\mathrm{init}}}|^2$, can be computed to be
\begin{align}
    \frac{U_q^2}{\vol^2} ~ \sin^2\left( \nicefrac{ \theta_{\kk1,\kk1+\qq} }{2}\right) \sin^2\left( \nicefrac{ \theta_{\kk2,\kk2-\qq} }{2}\right) ~ \delta_{\kk4,\kk2-\qq} \punc,
\end{align}
where we have defined $\qq = \kk3 - \kk1$, and where $\theta_{\kk1,\kk1+\qq}$ ($\theta_{\kk2,\kk2-\qq}$) is the angle between initial wavevector $\kk1$ ($\kk2$) and final wavevector $\kk1+\qq$ ($\kk2-\qq$). Using Fermi's Golden Rule in \eqnref{sup:FermiGoldenRule}, we can insert this into the expression for the relaxation time constant in \eqnref{sup:exprtaucc}, and convert the sums over momentum space to continuous integrals.
\begin{align}
\frac{1}{\tau_{\mathrm{cc}}} = & \frac{4 \pi}{\hbar} \iiint \frac{\dd{\kk1} \dd{\kk2} \dd{\qq}}{(2\pi)^6} \, U_q^2 \, \Phi^1(\ee1^-, \ee2^-, \ee3^+, \ee4^+) \nonumber \\[4pt]
\times ~ & \sin^2\left( \nicefrac{ \theta_{\kk1,\kk1+\qq} }{2}\right) \sin^2\left( \nicefrac{ \theta_{\kk2,\kk2-\qq} }{2}\right) \label{eq:sup:taunottransformed}\\
\times ~ & \, \delta(\ee1^- + \ee2^- - \e{\kk1+\qq}^+ - \e{\kk2-\qq}^+) ~~\mathclap{\Bigg/}~~\mathclap{\int}\, \frac{\dd{\kk1}}{(2\pi)^2}\,\Phi^0(\ee1^+, \ee1^-) \punc. \nonumber
\end{align}

\subsubsection{Defining Variable Transformations}
What follows is a series of variable transformations. As in the impurity case, we use $k_i\,\dd{k_i} = \frac{2m}{\hbar^2} \, \dd{\e{i}}\,\frac{1}{2}\,(1\pm\sqrt{\nicefrac{\ER}{\e{i}}})$ ($i=1,2$), and the denominator again becomes $\frac{1}{2\pi}\,\frac{2m}{\hbar^2}\,\beta^{-1}$. Next, we follow an approach by Lawrence and Wilkins~\cite{Lawrence73} to write the integrals over $\kk1$ and $\kk2$ in terms of the variables $\e1=\ee1^-$, $\e2=\ee2^-$, $\e{p}=\e{\kk1+\qq}^+$ and $\e{p'}=\e{\kk2-\qq}^+$.

First, we find an expression for the angle between $\kk1$ and $\qq$, which we call $\theta_{\kk1,\qq}$, as a function of the new variables. It is $\e1 = \nicefrac{\hbar^2}{2m} \, (k_1-\kR)^2$ and $\e{p} = \nicefrac{\hbar^2}{2m} \, (|\kk1+\qq|+\kR)^2$, and by writing $(\kk1+\qq)^2 = k_1^2 + q^2 + 2 k_1 q \cos \theta_{\kk1,\qq}$ we find that
\begin{align}
  \cos \theta_{\kk1,\qq} &=  \frac{|\kk1+\qq|^2 - k_1^2 - q^2}{2 k_1 q} \label{eq:sup:cosexpr} \\
  &= -\frac{\e1 - \e{p} + \e{q} + 2 \sqrt{\ER}(\sqrt{\e{p}} + \sqrt{\e1})}{2 \sqrt{\e{q}} (\sqrt{\e1} + \sqrt{\ER})}
\end{align}
and accordingly for the angle $\theta_{\kk2,\qq}$ between $\kk2$ and $\qq$ and variable $\e{p'}$. Therefore, we find the derivative of the angles with respect $\e{p}$ and $\e{p'}$ to be
\begin{align}
    \pdb{\theta_{\kk1,\qq}}{\e{p}} &= - \Omega(\e1,\e{p},\e{q}) \, \left(1 - \sqrt{\frac{\ER}{\e{p}}}\right) \punc, \\
    \pdb{\theta_{\kk2,\qq}}{\e{p'}} &= \hphantom- \Omega(\e2,\e{p'},\e{q}) \left(1 - \sqrt{\frac{\ER}{\e{p'}}}\right) \punc.
\end{align}
where $\Omega(\e1,\e{p},\e{q})$ is defined as
\begin{equation}
    \frac{1}{\sqrt{4 \e{q} \! \left(\sqrt{\e1} \!+\! \sqrt{\ER}\right)^2 \,\mathclap{-}\, \left(\e1 \!-\! \e{p} \!+\! \e{q} \!+\! 2 \sqrt{\ER}(\sqrt{\e{p}} \!+\! \sqrt{\e1})\right)^2}} \punc. \label{eq:sup:defomega}
\end{equation}
Using this result, we can find the Jacobian determinants for the transformation from $\kk1$ to $(\e1, \e{p})$ and $\kk2$ to $(\e2, \e{p'})$. We find that $k_1 \dd{k_1} \, \dd{\theta_{\kk1,\qq}}$ equals to
\begin{equation}
    \dd{\e1}\dd{\e{p}} \, \frac{2m}{\hbar^2} ~ \Omega(\e1,\e{p},\e{q}) \, \left(1 + \sqrt{\frac{\ER}{\e1}}\right) \left(1 - \sqrt{\frac{\ER}{\e{p}}}\right) \punc{,} \label{eq:sup:fulltransform}
\end{equation}
and the same expression for $k_2 \dd{k_2} \, \dd{\theta_{\kk2,\qq}}$ under exchange of $\e1 \mapsto \e2$ and $\e{p} \mapsto \e{p'}$. As was done previously, we drop the last two terms in the brackets as these small corrections account for the difference in the densities of states. We have included an extra factor of $2$ because the $\cos$ is only uniquely defined on the $[0,\pi]$ interval, and so accounts for only half of the integral that we want to calculate. The boundaries of the $\e{p}$ integral will be such that the momenta of $\kk1$ and $\qq$ either align or antialign, and so we find
\begin{align}
    \e{p}^{\nicefrac{\max}{\min}} &= \frac{\hbar^2}{2m} (\,|k_1 \pm q| + \kR)^2
\intertext{which can be rewritten as}
    \e{p}^{\max} &= ( \sqrt{\e1} + \sqrt{\e{q}} + 2\sqrt{\ER} )^2, \label{eq:sup:boundariesep}\\
    \e{p}^{\mathrlap{\min}\hphantom{\max}} &= \left\{\arr{ll}{
    ( \sqrt{\e1} - \sqrt{\e{q}} + 2\sqrt{\ER} )^2 \punc, &\mathrm{for}~\e{q} < \e1 \punc, \\
    ( \sqrt{\e{1}} - \sqrt{\e{q}} )^2 \punc, &\mathrm{for}~\e{q} > \e1 }\right.,\nonumber
\end{align}
and the respective result for $\e{p'}$ with $\e1 \mapsto \e2$. Finally, it is $q\,\dd{q} = \nicefrac{1}{2} \, (\nicefrac{2m}{\hbar^2}) \, \dd{\e{q}}$ with  $\e{q} = \nicefrac{\hbar^2 q^2}{2m}$. The integration over the remaining free angle results in an additional factor of $2\pi$. Using this and \eqnref{sup:fulltransform}, we can rewrite \eqnref{sup:taunottransformed} as
\begin{align}
\frac{1}{\tau_{\mathrm{cc}}} = & \frac{1}{2\pi\hbar} \left(\frac{2m}{\hbar^2}\right)^2 \! \beta \int_{\mathrlap{-\infty}}^{\mathrlap{+\infty}} \dd{\e1} \int_{\mathrlap{-\infty}}^{\mathrlap{+\infty}} \dd{\e2} \int_{\mathrlap{0}}^{\mathrlap{\infty}} \dd{\e{q}} \int_{\mathrlap{\e{p}^{\min}}}^{\mathrlap{\e{p}^{\max}}} ~\, \dd{\e{p}} \int_{\mathrlap{\e{p'}^{\min}}}^{\mathrlap{\e{p'}^{\max}}} ~\, \dd{\e{p'}} \nonumber\\
& \!\!\!\!\!\! \times U_{\e{q}}^2 \, \Phi^1(\e1,\e2,\e{p},\e{p'}) \, S(\e1,\e{p}) S(\e2,\e{p'}) \nonumber \\
& \!\!\!\!\!\! \times \Omega(\e1,\e{p},\e{q}) \Omega(\e2,\e{p'},\e{q}) \, \delta(\e1\!+\!\e2\!-\!\e{p}\!-\!\e{p'}), \label{eq:sup:tauagain}
\end{align}
where the transformed Coulomb potential, $U_{\e{q}}$, and spinor overlap matrix elements $S(\e{1},\e{p})$ and $S(\e{2},\e{p'})$ will be provided in the next section.

\subsubsection{Applying Variable Transformations to Integrand}
Having defined those variable transformations and having calculated their Jacobian determinants, we will proceed by applying the transformation to the integrand.

The Coulomb potential can easily be rewritten as
\begin{equation}
    U_{\e{q}} = \left(\frac{2m}{\hbar^2}\right)^{-\nicefrac{1}{2}} \!\!\! \frac{2\pi e_0^2}{\sqrt{\e{q}} + \sqrt{\e{S}}}\punc,
\end{equation}
where we have defined $\e{S} = \nicefrac{\hbar^2 \kS^2}{2m}$.

We can relate the angle between $\kk1$ and $\kk1+\qq$ ($\kk2$ and $\kk2-\qq$) to the angle between $\kk1$ ($\kk2$) and $\qq$,
\begin{align}
    \cos\left(\theta_{\kk1,\kk1+\qq}\right) &= \frac{k_1 + q \cos\left(\theta_{\kk1,\qq}\right)}{|\kk1+\qq|} \punc{,} \\ \cos\left(\theta_{\kk2,\kk2-\qq}\right) & = \frac{k_2 - q \cos\left(\theta_{\kk2,\qq}\right)}{|\kk2-\qq|} \punc{,}
\end{align}
and with $\sin^2(\nicefrac{\theta}{2}) = \nicefrac{1}{2} \, (1-\cos(\theta))$, we find
\begin{align}
    \sin^2\left(\nicefrac{\theta_{\kk1,\kk1+\qq}}{2}\right) & \!\!\! \overset{\hphantom{\eqref{eq:sup:cosexpr}}}{=} \!\! \frac{1}{2} \frac{|\kk1+\qq| - k_1 - q \cos\left(\theta_{\kk1,\qq}\right)}{|\kk1+\qq|} \\
    & \!\!\! \overset{\eqref{eq:sup:cosexpr}}{=} \!\! \frac{1}{4} \frac{q^2 - (k_1 - |\kk1+\qq|)^2}{k_1 |\kk1+\qq|} \punc, \\
    \sin^2\left(\nicefrac{\theta_{\kk2,\kk2-\qq}}{2}\right) & \!\!\! \overset{\hphantom{\eqref{eq:sup:cosexpr}}}{=} \!\! \frac{1}{2} \frac{|\kk2-\qq| - k_2 + q \cos\left(\theta_{\kk2,\qq}\right)}{|\kk2-\qq|} \\
    & \!\!\! \overset{\hphantom{\eqref{eq:sup:cosexpr}}}{=} \!\! \frac{1}{4} \frac{q^2 - (k_2 - |\kk2-\qq|)^2}{k_2 |\kk2-\qq|} \punc.
\end{align}
Therefore, we can define the spinor overlap function $S(\e1,\e{p},\e{q})$ used in \eqnref{sup:tauagain} as
\begin{align}
    \frac{1}{4} \frac{\e{q} - (\sqrt{\e1} - \sqrt{\e{p}} + 2\sqrt{\ER})^2}{(\sqrt{\e1} + \sqrt{\ER})(\sqrt{\e{p}} - \sqrt{\ER})} \punc.
\end{align}

\subsubsection{Computing the Integral}
In this section, we will proceed by computing the integral in \eqnref{sup:tauagain}. This will yield an expression for the inter-carrier scattering relaxation timescale $\tau_{\mathrm{cc}}$.

We define $\e{\Delta} = \e{p} - \e1$ and $\e{\Delta'} = \e{p'} - \e2$ and write \eqnref{sup:tauagain} as
\begin{align}
\frac{1}{\tau_{\mathrm{cc}}} = & \frac{\beta}{\tau_{cc,0}} \! \int_{\mathrlap{-\infty}}^{\mathrlap{+\infty}} \dd{\e1} \int_{\mathrlap{-\infty}}^{\mathrlap{+\infty}} \dd{\e2} \int_{\mathrlap{0}}^{\mathrlap{\infty}} \dd{\e{q}} \int_{\mathrlap{\e{\Delta}^{\min}}}^{\mathrlap{\e{\Delta}^{\max}}} ~ \dd{\e{\Delta}} \int_{\mathrlap{\e{\Delta}^{\min}}}^{\mathrlap{\e{\Delta}^{\max}}} ~ \dd{\e{\Delta}} \, u_{\e{q}}^2 \\
& \!\!\!\!\!\! \times \Phi^1(\e1,\e2,\e1\!+\!\e{\Delta},\e2\!+\!\e{\Delta'}) \, S(\e1,\e1\!+\!\e{\Delta}) S(\e2,\e2\!+\!\e{\Delta'}) \nonumber \\
& \!\!\!\!\!\! \times \Omega(\e1,\e1+\e{\Delta},\e{q}) \Omega(\e2,\e2+\e{\Delta'},\e{q}) \, \delta(\e{\Delta}\!+\!\e{\Delta'}) \punc, \nonumber
\end{align}
where $u_{\e{q}} = (\sqrt{\e{q}} + \sqrt{\e{S}})^{-1}$ and
\begin{equation}
    \frac{1}{\tau_{cc,0}} = \frac{(2\pi e_0^2)^2}{2\pi\hbar} \, \frac{2m}{\hbar^2} = \frac{1}{8\pi} \frac{m}{\hbar^3} \frac{e^4}{\epsilon_0^2 \, \kappa^2} \punc.
    \label{eq:sup:deftau0}
\end{equation}
The boundaries of the integration over $\e{\Delta}$ can be deduced from \eqnref{sup:boundariesep} and are given by
\begin{align}
    \e{\Delta}^{\nicefrac{\max}{\min}} &=
    (2\sqrt{\ER} \pm \sqrt{\e{q}})^2 + 2 \sqrt{\e1} (2\sqrt{\ER} \pm \sqrt{\e{q}}) \punc, \\
\intertext{for $\e{q} < \e1$ and}
    \e{\Delta}^{\max} &=
    (2\sqrt{\ER} + \sqrt{\e{q}})^2 + 2 \sqrt{\e1} (2\sqrt{\ER} + \sqrt{\e{q}}) \punc, \\
    \e{\Delta}^{\mathrlap{\min}\hphantom{\max}} &= \e{q} - 2 \sqrt{\e1\e{q}} \punc,
\end{align}
for $\e{q} > \e1$, (and accordingly for $\e{\Delta'}$).

In order for the delta function to give a contribution, it must be $\e{\Delta}^{\max} + \e{\Delta'}^{\max} > 0$ and $\e{\Delta}^{\min} + \e{\Delta'}^{\min} < 0$. The first condition is trivial, and can be omitted. In the case $\e{q} < \e1$, the second condition requires that
\begin{equation}
    \left(2\sqrt{\ER} \!-\! \sqrt{\e{q}}\right)^2 \!\!+\! \left(\sqrt{\e1}\!+\!\sqrt{\e2}\right) \left(2\sqrt{\ER} \!-\! \sqrt{\e{q}} \right) < 0 \punc.
\end{equation}
This necessitates that $\sqrt{\e{q}} > 2\sqrt{\ER}$, which is just the observation that $q>2\kR$ that we discuss in detail in the main text. Under this condition, the inequality is inverted when dividing by $\left( 2\sqrt{\ER} - \sqrt{\e{q}} \right)$, and we find
\begin{equation}
    \left(2\sqrt{\ER} - \sqrt{\e{q}}\right) + \left(\sqrt{\e1}+\sqrt{\e2}\right) > 0 \punc,
\end{equation}
which is always true for the case $\e{q} < \e1$ that we started with. In the case of $\e{q} > \e1$ (in which case we also assert that $\e{q} > \e2$), the condition $\e{\Delta}^{\min} + \e{\Delta'}^{\min} < 0$ requires that
\begin{equation}
    \sqrt{\e{q}} < \sqrt{\e1} + \sqrt{\e2} \punc.
\end{equation}

We can now execute the integration over $\e{\Delta'}$, which is equivalent to setting $\e{\Delta'} = -\e{\Delta}$ due to the $\delta$ function. The integration over $\e{\Delta}$ will then run from $\e{\min} = \max\left(\e{\Delta}^{\min}, -\e{\Delta'}^{\max}\right)$ to $\e{\max} = \min\left(\e{\Delta}^{\max}, -\e{\Delta'}^{\min}\right)$. To find the $\min$ or $\max$, we check whether $\e{\Delta}^{\nicefrac{\min}{\max}} + \e{\Delta'}^{\nicefrac{\max}{\min}} > 0$, which in the case of $\e{q}<\e1$ is equivalent to
\begin{equation}
    4\ER+\e{q} + 2\sqrt{\ER} (\sqrt{\e1}\!+\!\sqrt{\e2}) \mp \sqrt{\e{q}} (\sqrt{\e1}\!-\!\sqrt{\e2}) > 0 \punc.
\end{equation}
Because $|\sqrt{\e1}\!-\!\sqrt{\e2}| \approx \sqrt{\kB T} \ll \sqrt{\ER} < \sqrt{\e{q}}$, this is always true. Furthermore, in the case $\e{q}>\e1$ the above condition is equivalent to
\begin{equation}
    (2\sqrt{\ER} \!+\! \sqrt{\e{q}})^2 \!+\! \e{q} + 2 \sqrt{\e{\nicefrac{2}{1}} \ER} \!\mp\! 2\sqrt{\e{q}} (\sqrt{\e1} \!-\! \sqrt{\e2}) > 0 \punc,
\end{equation}
which is satisfied again because $|\sqrt{\e1}\!-\!\sqrt{\e2}| \ll \sqrt{\e{q}}$. Therefore, $\e{\min} = \e{\Delta}^{\min}$ and $\e{\max} = -\e{\Delta'}^{\min}$. Using these results, we can now write
\begin{align}
\frac{1}{\tau_{\mathrm{cc}}} = & \frac{\beta}{\tau_{\mathrm{cc},0}} \int\!\dd{\e1}\!\int\!\dd{\e2}\!\int_{4\ER}^{\mathrlap{(\sqrt{\e1}+\sqrt{\e2})^2}} \dd{\e{q}} ~~~~ \int_{\e{\min}}^{\e{\max}} \dd{\e\Delta} ~ u_{\e{q}}^2 \\
& \!\!\!\!\!\!\!\!\!\!\!\! \times \Phi^1(\e1,\e2,\e1\!+\!\e\Delta,\e2\!-\!\e\Delta) \, S(\e1,\e1\!+\!\e\Delta) S(\e2,\e2\!-\!\e\Delta) \nonumber \\
& \!\!\!\!\!\!\!\!\!\!\!\! \times \Omega(\e1,\e1+\e\Delta,\e{q}) \Omega(\e2,\e2-\e\Delta,\e{q}) \punc. \nonumber
\end{align}

\begin{figure}[!t]
    \centering
    \includegraphics[width=\linewidth]{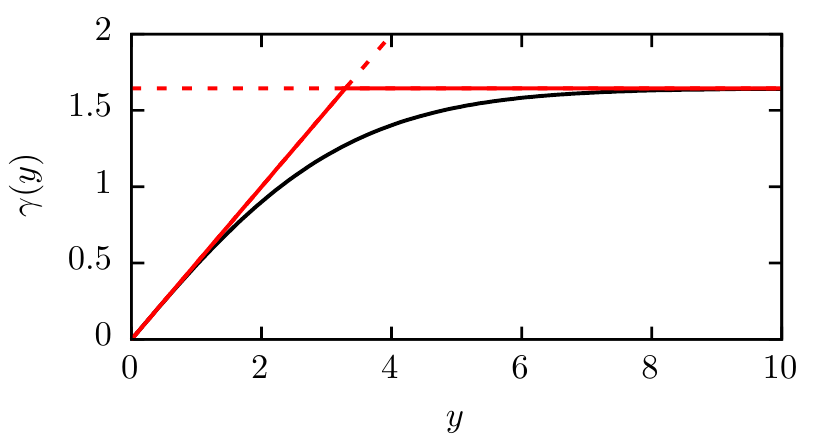}
    \caption{\label{fig:figS01}
    Approximation of $\gamma(y)$ as defined in \eqnref{sup:gamma_approximated}.}
\end{figure}

So far, all manipulations have been exact. We will continue with a number of approximations to be able to find a closed expression for the relaxation time. The spinor overlap function $S(\e1, \e1+\e\Delta, \e q)$ takes the form
\begin{align}
    \frac{1}{4} \frac{\e{q} - (\sqrt{\e1} - \sqrt{\e1+\e\Delta} + 2\sqrt{\ER})^2}{(\sqrt{\e1} + \sqrt{\ER})(\sqrt{\e1+\e\Delta} - \sqrt{\ER})} \punc.
\end{align}
$\e\Delta$ is of order $\kB T$ and therefore much smaller than $\e1 \approx \mu$ and $\ER$. We therefore expand the denominator in lowest order in $\e\Delta$ and drop its dependence in the numerator, which gives
\begin{align}
    \frac{1}{4} \frac{\sqrt{\e1} (\e{q} - 4 \ER) + 2 \e\Delta \sqrt{\ER}}{\sqrt{\e1} (\e1 - \ER)} \punc.
\end{align}
Similarly, we approximate $\Omega(\e1,\e1+\e\Delta,\e{q})$ (defined in \eqnref{sup:defomega}) as
\begin{equation}
    \frac{1}{\sqrt{4 \e{q} \! \left(\sqrt{\e1} \!+\! \sqrt{\ER}\right)^2 \,\mathclap{-}\, \left(-\e\Delta \!+\! \e{q} \!+\! 4 \sqrt{\e1\ER}\right)^2}} \punc.
\end{equation}
Next, we note that $\e1$ and $\e2$ are approximately equal to $\mu$ and variations around this value are on the order of $\kB T$. We can therefore neglect these variations and replace $\e1$ and $\e2$ by $\mu$ everywhere except for in the Fermi distributions in $\Phi^1$. Consequently, we can perform the integrations over $\e1$ and $\e2$. We use the following expression for $\Phi^1(\e1, \e2, \e1+\e\Delta, \e2-\e\Delta)$,
\begin{align}
       &\f{\e1+\e\Delta} \f{\e2-\e\Delta} [1-\f{\e1}] [1-\f{\e2}] \\
       ~\times& \Big( [1-\f{\e1+\e\Delta}] + [1-\f{\e2-\e\Delta}] + \f{\e1} + \f{\e2}\Big) \nonumber \\
      +&\f{\e1} \f{\e2} [1-\f{\e1+\e\Delta}] [1-\f{\e2-\e\Delta}] \nonumber\\
       ~\times& \Big( \f{\e1+\e\Delta} + \f{\e2-\e\Delta} + [1-\f{\e1}] + [1-\f{\e2}] \Big) \punc, \nonumber
\end{align}
where the second term gives the same contribution as the first, as can be seen by renaming $\e1+\e\Delta \mapsto \e2$ and $\e2-\e\Delta \mapsto \e1$. Finally, a straight-forward integration over $\e1$ and $\e2$ gives,
\begin{align}
    & \int\!\!\e1\!\!\int\!\!\e2\,\Phi^1(\e1,\e2, \e1\!+\!\e\Delta, \e2\!-\!\e\Delta) = \frac{4 \, \e{\Delta}^2 \, e^{\beta\e{\Delta}}}{(e^{\beta\e{\Delta}}-1)^2} \punc.
\end{align}
This gives
\begin{align}
\frac{1}{\tau_{\mathrm{cc}}} = \, &\frac{(\kB T)^2}{\tau_{\mathrm{cc},0}} \!\! \int_{4\ER}^{4\mu} \!\! \dd{\e{q}} \, u_{\e{q}}^2 \int_{-\e{\max}}^{\e{\max}} \!\! \beta \, \dd{\e\Delta} \\
& \!\!\!\!\!\! \times \frac{4 (\beta\e{\Delta})^2 e^{\beta\e{\Delta}}}{(e^{\beta\e{\Delta}}-1)^2} \frac{1}{16} \frac{\mu (\e{q}-4\ER)^2 - 4 \e\Delta^2 \ER}{\mu (\mu - \ER)^2} \nonumber \\
& \!\!\!\!\!\! \times \frac{1}{\sqrt{4 \e{q} (\sqrt{\mu} + \sqrt{\ER})^2 - (\e{q} + 4\sqrt{\mu \ER} + \e\Delta)^2}} \nonumber \\
& \!\!\!\!\!\! \times \frac{1}{\sqrt{4 \e{q} (\sqrt{\mu} + \sqrt{\ER})^2 - (\e{q} + 4\sqrt{\mu \ER} - \e\Delta)^2}} \punc. \nonumber
\end{align}
We use the lowest order expansion in $\e\Delta$ and use the substitution $x = \beta\e{\Delta}$ to find
\begin{align}
\frac{1}{\tau_{\mathrm{cc}}} = \, &\frac{1}{\tau_{cc,0}} \frac{(\kB T)^2}{(\mu-\ER)^2} \! \int_{\mathrlap{4\ER}}^{\mathrlap{4\mu}} ~ \dd{\e{q}} \, \frac{\e{q}-4\ER}{\e{q}(4\mu-\e{q})} \, \gamma(\beta\e{\max}) \punc,
\end{align}
where we've set $\e{S} = 0$ (because its effect on screening can be neglected), and where we have defined
\begin{equation}
 \gamma(y) = \frac{1}{2}\int_0^y \dd{x} ~ \frac{x^2 e^x}{(e^x-1)^2} \punc.
\end{equation}
Where $\epsilon_{\max} = \sqrt{\e{q}}(\sqrt{4\mu}-\sqrt{\e{q}}) \gg \kB T$, we can replace $\gamma(\beta\epsilon_{\max})$ with $\gamma(\infty) = \frac{\pi^2}{6}$. This holds everywhere except for $\e{q}$ close to $4\mu$, which is the upper limit of the $\e{q}$ integral and also where the rest of the integrand is logarithmically divergent (due to the $(4\mu-\e{q})^{-1}$ contribution). We therefore have to split the $\e{q}$ integration up into one part close to $\e{q}=4\mu$ and one part further away from this point. As shown in \figref{figS01}, we can approximate $\gamma(y)$ as
\begin{equation}
    \gamma(y) = \left\{ \arr{cl}{\frac{\pi^2}{6} \punc, & \mathrm{if}~y  >   \frac{\pi^2}{3} \punc, \\
                                 \frac{y}{2} \punc,               & \mathrm{if}~y \leq \frac{\pi^2}{3} \punc.} \right. \label{eq:sup:gamma_approximated}
\end{equation}
We solve $\beta\epsilon_{\max} = \frac{\pi^2}{6}$ for $\e{q}$ to find
\begin{align}
    \e{q} &= 2\mu \, \left(1 - \frac{\pi^2}{6} \frac{\kB T}{\mu} + \sqrt{1-\frac{\pi^2}{3} \frac{\kB T}{\mu}} \right) \\
    &\approx 4\mu \left(1-\frac{\pi^2}{6} \frac{\kB T}{\mu} \right) =: \e{q}^* \punc.
\end{align}
We can therefore write
\begin{align}
\frac{1}{\tau_{\mathrm{cc}}}
  =~& \int_{\mathrlap{4\ER}}^{\mathrlap{4\mu}} ~ \dd{\e{q}} \, \frac{\e{q}-4\ER}{\e{q}(4\mu-\e{q})} \, \gamma(\beta\e{\max}) \\
  =~& \frac{\pi^2}{6}\int_{\mathrlap{4\ER}}^{\mathrlap{\e{q}^*}} ~ \dd{\e{q}} \, \frac{\e{q}-4\ER}{\e{q}(4\mu-\e{q})} \\\nonumber
  +~& \frac{1}{2} \int_{\mathrlap{\e{q}^*}}^{\mathrlap{4\mu}} ~ \dd{\e{q}} \, \frac{\e{q}-4\ER}{\sqrt{\e{q}}(\sqrt{4\mu}+\sqrt{\e{q}})} \nonumber\punc,
\intertext{and since $\e{q}^* \approx 4\mu$ we can set $\e{q} = 4\mu$ in the second integral, which results in}
\frac{1}{\tau_{\mathrm{cc}}}
  =~& \frac{\pi^2}{6}\!\left[ - \frac{\ER}{\mu} \log(\e{q}) - \frac{(\mu\!-\!\ER) \log(4\mu\!-\!\e{q})}{\mu} \right]_{\e{q}=4\ER}^{\e{q}=\e{q}^*} \nonumber \\
 +~~& \frac{4\mu}{2} \, \frac{\pi^2}{6} \frac{\kB T}{\mu} \frac{4\mu-4\ER}{2\sqrt{\mu}\,(\sqrt{4\mu}+\sqrt{4\mu})} \\
  \approx~& \frac{\pi^2}{6} \Bigg(1-\frac{\ER}{\mu}\left(1-\log\left(\frac{\ER}{\mu}\right)\right) \nonumber \\
  & \, - \left(1-\frac{\ER}{\mu}\right) \, \log\left(\frac{\pi^2}{6}\frac{\kB T}{\mu-\ER}\right) \Bigg) \punc,
\end{align}
which we define as $\rho\left(\frac{\ER}{\mu}, \frac{\kB T}{\mu-\ER}\right)$. We can therefore write
\begin{equation}
\frac{1}{\tau_{\mathrm{cc}}} = \frac{1}{\tau_{\mathrm{cc},0}} \, \frac{(\kB T)^2}{(\mu-\ER)^2} \times \rho\left(\frac{\ER}{\mu}, \frac{\kB T}{\mu-\ER}\right) \punc.
\end{equation}
Using \eqnref{sup:deftau0} and $\kappa = 10$, we find $\tau_{\mathrm{cc},0} = \SI{E-11}{\second}$. For thin films of GeTe with $\mu - \ER = \SI{0.1}{\electronvolt}$~\cite{Liebmann16, Narayan16} and $\frac{\ER}{\mu} \approx 0.5$, which gives $\rho\left(\frac{\ER}{\mu}, \frac{\kB T}{\mu-\ER}\right) \approx 10$ and therefore yields $\tau_{\mathrm{cc}} \approx \SI{1}{\micro\second}$ at $T=\SI{100}{\milli\kelvin}$ as reported in \secref{coulomb}.

\section{The role of Spin Flips}
\label{sec:spinflips}
While the phonon-mediated transitions are prohibited independently of spin due to energy-momentum conservation, we made explicit use of the anti-alignment of spin eigenstate for low-$q$ inter-band transitions mediated by the Coulomb interaction in \secref{coulomb} and assumed the absence of spin-flip processes. It is however well known that the spin of a carrier can be flipped through various different mechanisms, and it is therefore worthwhile asking whether any of these can enhance the inter-band scattering rate and ultimately invalidate the arguments presented in \secref{coulomb}.

The spin of a carrier can either be flipped through interaction with magnetic impurities, which can be disregarded in high-purity samples, or alternatively through strong spin-orbit coupling, which is a crucial ingredient of our work. One may therefore ask whether the spin-orbit coupling of the Rashba system can induce spin flips that will lead to fast inter-band equilibration.

First we note that it is important not to confuse the spin relaxation timescale with the likelihood for spin-reversing processes to occur. We expect spin relaxation to continue to occur on very short timescales in our system, as spin is not a conserved quantity in each individual Rashba band and scattering events that conserve the chiral index will occur at a high rate. To relax the non-equilibrium carrier configuration that we study, spin-reversing inter-band scattering processes are required and there is no direct relation between their rate and the spin relaxation timescale.

Next, we note that while spin-orbit coupling by itself leads to spin mixing of electronic eigenstates, no spin flips can occur without a scattering process that mediates it. We have identified all relevant scattering processes for the relaxation of the depicted non-equilibrium state in \secref{model}, and we would now have to continue by examining the effect of the spin-orbit coupling onto those mechanisms.

As explained in \secref{phonons}, phonon-mediated transitions between the Rashba bands are suppressed due to energy-momentum conservation completely independently of the relative spin alignments. If we assume a high-purity sample and ignore carrier-impurity scattering for the moment, then inter-carrier scattering is the only way of inducing transitions between states of opposite chirality. We therefore have to check how the addition of spin-orbit coupling affects these transitions.

It is important to understand that the momentum-dependent spin mixing of the Rashba coupling has already been taken into account explicitly in our calculations in \secref{coulomb} when we included the spinor part of the wavefunctions in the matrix element. It will therefore only be necessary to probe the effect of momentum-dependent spin mixing to other Bloch bands, and not between the two Rashba bands.

Phonon-mediated spin-orbit induced spin flips can be described using the language of the Elliot-Yafet mechanism~\cite{Baral16}, where the momentum-dependent spin mixing to other Bloch bands is referred to as the Elliot contribution \cite{PhysRev.96.266}, and the effect of the phonon-modulated spin-orbit interaction referred to as the Overhauser contribution \cite{PhysRev.89.689}. For the inter-carrier scattering, the Overhauser part is absent, and we will have to focus on the Elliot contribution only. We will now show that this vanishes up to first order in $q$. As we explain in \secref{coulomb}, inter-carrier scattering is dominated by low-$q$ transitions and therefore incompatible with this $q$ dependence of the Elliot contribution.

We determine the scattering rate for spin-reversing inter-band processes by adding the spin mixing to another Bloch band to the transition matrix element in \eqnref{FermiGoldenRule}. This is done by adding the product of the lattice-periodic part $u_{\mu\kk{}}$ of the Bloch functions $\varphi_{\mu\kk{}} (\rr{}) = \exp(i\kk{}\rr{}) \, u_{\mu\kk{}} (\rr{})$ of initial and final state to the real-space integral in the expectation value. In momentum space, this results in multiplying the outcome of the result from \secref{coulomb} with the Fourier transform of the product of the initial and final $u_{\mu\kk{}}$,
\begin{equation}
    \int\ddd{r}{2}\,\exp\left(i \qq \rr{} \right)\,u^*_{\mu\kk{}} u_{\mu'\kk{}+\qq} \punc.
\end{equation}
Note that the labels $\mu$ and $\mu'$ refer to both the Bloch band index and the pseudospin index that labels the spin degree of freedom. We can expand this integral around $q=0$ and examine the leading-order terms. The zeroth order term is the overlap of the two wavefunctions with same momentum and opposite spin index, and is zero as the $u_{\mu\kk{}}$ are orthogonal. The term in first order integrates over the $\rr{}$ vector,
\begin{equation}
    i\qq\int\ddd{r}{2}\,u^*_{\mu\kk{}}\,\rr{}\,u_{\mu'\kk{}+\qq} \punc.
\end{equation}
This term however must vanish because it can only exist in a system that breaks inversion symmetry \footnote{Note that we refer to the symmetry within the 2D plane, while obviously the Rashba interaction breaks the inversion symmetry in the direction perpendicular to the plane}. Therefore, we can conclude that the lowest-order term in the small-$q$ expansion is proportional to $q^2$.

This is the main reason why typically phonons are considered as the main source for spin-orbit induced spin-flips, as they are able to provide sufficient change of momentum to be compatible with the vanishing of the momentum-dependent spin mixing for small momentum transfer\footnote{Check Ref.~\cite{Baral16} for a detailed discussion of the low-$q$ expansion of the Elliot-Yafet mechanism.}. In other words, we can expect the spin mixing to have a similar effect in inhibiting inter-carrier transitions as the opposing spin alignments of the Rashba spin texture had in \secref{coulomb}. This leaves us to conclude this section by summarising that spin-orbit coupling (Rashba coupling and to other Bloch bands) is ineffective in enabling inter-band transitions to opposite spin states.

\bibliography{lit}

\begin{thebibliography}{27}%
\makeatletter
\providecommand \@ifxundefined [1]{%
 \@ifx{#1\undefined}
}%
\providecommand \@ifnum [1]{%
 \ifnum #1\expandafter \@firstoftwo
 \else \expandafter \@secondoftwo
 \fi
}%
\providecommand \@ifx [1]{%
 \ifx #1\expandafter \@firstoftwo
 \else \expandafter \@secondoftwo
 \fi
}%
\providecommand \natexlab [1]{#1}%
\providecommand \enquote  [1]{``#1''}%
\providecommand \bibnamefont  [1]{#1}%
\providecommand \bibfnamefont [1]{#1}%
\providecommand \citenamefont [1]{#1}%
\providecommand \href@noop [0]{\@secondoftwo}%
\providecommand \href [0]{\begingroup \@sanitize@url \@href}%
\providecommand \@href[1]{\@@startlink{#1}\@@href}%
\providecommand \@@href[1]{\endgroup#1\@@endlink}%
\providecommand \@sanitize@url [0]{\catcode `\\12\catcode `\$12\catcode
  `\&12\catcode `\#12\catcode `\^12\catcode `\_12\catcode `\%12\relax}%
\providecommand \@@startlink[1]{}%
\providecommand \@@endlink[0]{}%
\providecommand \url  [0]{\begingroup\@sanitize@url \@url }%
\providecommand \@url [1]{\endgroup\@href {#1}{\urlprefix }}%
\providecommand \urlprefix  [0]{URL }%
\providecommand \Eprint [0]{\href }%
\providecommand \doibase [0]{http://dx.doi.org/}%
\providecommand \selectlanguage [0]{\@gobble}%
\providecommand \bibinfo  [0]{\@secondoftwo}%
\providecommand \bibfield  [0]{\@secondoftwo}%
\providecommand \translation [1]{[#1]}%
\providecommand \BibitemOpen [0]{}%
\providecommand \bibitemStop [0]{}%
\providecommand \bibitemNoStop [0]{.\EOS\space}%
\providecommand \EOS [0]{\spacefactor3000\relax}%
\providecommand \BibitemShut  [1]{\csname bibitem#1\endcsname}%
\let\auto@bib@innerbib\@empty
\bibitem [{\citenamefont {Wolf}\ \emph {et~al.}(2001)\citenamefont {Wolf},
  \citenamefont {Awschalom}, \citenamefont {Buhrman}, \citenamefont {Daughton},
  \citenamefont {von Moln{\'a}r}, \citenamefont {Roukes}, \citenamefont
  {Chtchelkanova},\ and\ \citenamefont {Treger}}]{Wolf01}%
  \BibitemOpen
  \bibfield  {author} {\bibinfo {author} {\bibfnamefont {S.~A.}\ \bibnamefont
  {Wolf}}, \bibinfo {author} {\bibfnamefont {D.~D.}\ \bibnamefont {Awschalom}},
  \bibinfo {author} {\bibfnamefont {R.~A.}\ \bibnamefont {Buhrman}}, \bibinfo
  {author} {\bibfnamefont {J.~M.}\ \bibnamefont {Daughton}}, \bibinfo {author}
  {\bibfnamefont {S.}~\bibnamefont {von Moln{\'a}r}}, \bibinfo {author}
  {\bibfnamefont {M.~L.}\ \bibnamefont {Roukes}}, \bibinfo {author}
  {\bibfnamefont {A.~Y.}\ \bibnamefont {Chtchelkanova}}, \ and\ \bibinfo
  {author} {\bibfnamefont {D.~M.}\ \bibnamefont {Treger}},\ }\href {\doibase
  10.1126/science.1065389} {\bibfield  {journal} {\bibinfo  {journal}
  {Science}\ }\textbf {\bibinfo {volume} {294}},\ \bibinfo {pages} {1488}
  (\bibinfo {year} {2001})}\BibitemShut {NoStop}%
\bibitem [{\citenamefont {Wu}\ \emph {et~al.}(2010)\citenamefont {Wu},
  \citenamefont {Jiang},\ and\ \citenamefont {Weng}}]{wu_spin_2010}%
  \BibitemOpen
  \bibfield  {author} {\bibinfo {author} {\bibfnamefont {M.~W.}\ \bibnamefont
  {Wu}}, \bibinfo {author} {\bibfnamefont {J.~H.}\ \bibnamefont {Jiang}}, \
  and\ \bibinfo {author} {\bibfnamefont {M.~Q.}\ \bibnamefont {Weng}},\ }\href
  {\doibase 10.1016/j.physrep.2010.04.002} {\bibfield  {journal} {\bibinfo
  {journal} {Physics Reports}\ }\textbf {\bibinfo {volume} {493}},\ \bibinfo
  {pages} {61} (\bibinfo {year} {2010})}\BibitemShut {NoStop}%
\bibitem [{\citenamefont {{\v Z}uti{\'c}}\ \emph
  {et~al.}(2004{\natexlab{a}})\citenamefont {{\v Z}uti{\'c}}, \citenamefont
  {Fabian},\ and\ \citenamefont {{Das Sarma}}}]{zutic_spintronics_2004}%
  \BibitemOpen
  \bibfield  {author} {\bibinfo {author} {\bibfnamefont {I.}~\bibnamefont {{\v
  Z}uti{\'c}}}, \bibinfo {author} {\bibfnamefont {J.}~\bibnamefont {Fabian}}, \
  and\ \bibinfo {author} {\bibfnamefont {S.}~\bibnamefont {{Das Sarma}}},\
  }\href {\doibase 10.1103/RevModPhys.76.323} {\bibfield  {journal} {\bibinfo
  {journal} {Reviews of Modern Physics}\ }\textbf {\bibinfo {volume} {76}},\
  \bibinfo {pages} {323} (\bibinfo {year} {2004}{\natexlab{a}})}\BibitemShut
  {NoStop}%
\bibitem [{\citenamefont {Bihlmayer}\ \emph {et~al.}(2015)\citenamefont
  {Bihlmayer}, \citenamefont {Rader},\ and\ \citenamefont
  {Winkler}}]{Bihlmayer15}%
  \BibitemOpen
  \bibfield  {author} {\bibinfo {author} {\bibfnamefont {G.}~\bibnamefont
  {Bihlmayer}}, \bibinfo {author} {\bibfnamefont {O.}~\bibnamefont {Rader}}, \
  and\ \bibinfo {author} {\bibfnamefont {R.}~\bibnamefont {Winkler}},\ }\href
  {http://stacks.iop.org/1367-2630/17/i=5/a=050202} {\bibfield  {journal}
  {\bibinfo  {journal} {New Journal of Physics}\ }\textbf {\bibinfo {volume}
  {17}},\ \bibinfo {pages} {050202} (\bibinfo {year} {2015})}\BibitemShut
  {NoStop}%
\bibitem [{\citenamefont {{\v Z}uti{\'c}}\ \emph
  {et~al.}(2004{\natexlab{b}})\citenamefont {{\v Z}uti{\'c}}, \citenamefont
  {Fabian},\ and\ \citenamefont {Sarma}}]{Zutic04}%
  \BibitemOpen
  \bibfield  {author} {\bibinfo {author} {\bibfnamefont {I.}~\bibnamefont {{\v
  Z}uti{\'c}}}, \bibinfo {author} {\bibfnamefont {J.}~\bibnamefont {Fabian}}, \
  and\ \bibinfo {author} {\bibfnamefont {S.~D.}\ \bibnamefont {Sarma}},\
  }\href@noop {} {\bibfield  {journal} {\bibinfo  {journal} {Rev. Mod. Phys.}\
  }\textbf {\bibinfo {volume} {76}},\ \bibinfo {pages} {323} (\bibinfo {year}
  {2004}{\natexlab{b}})}\BibitemShut {NoStop}%
\bibitem [{\citenamefont {Nitta}\ \emph {et~al.}(1997)\citenamefont {Nitta},
  \citenamefont {Akazaki}, \citenamefont {Takayanagi},\ and\ \citenamefont
  {Enoki}}]{Nitta97}%
  \BibitemOpen
  \bibfield  {author} {\bibinfo {author} {\bibfnamefont {J.}~\bibnamefont
  {Nitta}}, \bibinfo {author} {\bibfnamefont {T.}~\bibnamefont {Akazaki}},
  \bibinfo {author} {\bibfnamefont {H.}~\bibnamefont {Takayanagi}}, \ and\
  \bibinfo {author} {\bibfnamefont {T.}~\bibnamefont {Enoki}},\ }\href
  {\doibase 10.1103/PhysRevLett.78.1335} {\bibfield  {journal} {\bibinfo
  {journal} {Phys. Rev. Lett.}\ }\textbf {\bibinfo {volume} {78}},\ \bibinfo
  {pages} {1335} (\bibinfo {year} {1997})}\BibitemShut {NoStop}%
\bibitem [{\citenamefont {Miller}\ \emph {et~al.}(2003)\citenamefont {Miller},
  \citenamefont {Zumb{\"u}hl}, \citenamefont {Marcus}, \citenamefont
  {Lyanda-Geller}, \citenamefont {Goldhaber-Gordon}, \citenamefont {Campman},\
  and\ \citenamefont {Gossard}}]{Miller03}%
  \BibitemOpen
  \bibfield  {author} {\bibinfo {author} {\bibfnamefont {J.~B.}\ \bibnamefont
  {Miller}}, \bibinfo {author} {\bibfnamefont {D.~M.}\ \bibnamefont
  {Zumb{\"u}hl}}, \bibinfo {author} {\bibfnamefont {C.~M.}\ \bibnamefont
  {Marcus}}, \bibinfo {author} {\bibfnamefont {Y.~B.}\ \bibnamefont
  {Lyanda-Geller}}, \bibinfo {author} {\bibfnamefont {D.}~\bibnamefont
  {Goldhaber-Gordon}}, \bibinfo {author} {\bibfnamefont {K.}~\bibnamefont
  {Campman}}, \ and\ \bibinfo {author} {\bibfnamefont {A.~C.}\ \bibnamefont
  {Gossard}},\ }\href {\doibase 10.1103/PhysRevLett.90.076807} {\bibfield
  {journal} {\bibinfo  {journal} {Physical Review Letters}\ }\textbf {\bibinfo
  {volume} {90}},\ \bibinfo {pages} {076807} (\bibinfo {year}
  {2003})}\BibitemShut {NoStop}%
\bibitem [{\citenamefont {Eschbach}\ \emph {et~al.}(2015)\citenamefont
  {Eschbach}, \citenamefont {M{\l}y{\'n}czak}, \citenamefont {Kellner},
  \citenamefont {Kampmeier}, \citenamefont {Lanius}, \citenamefont {Neumann},
  \citenamefont {Weyrich}, \citenamefont {Gehlmann}, \citenamefont
  {Gospodari{\v c}}, \citenamefont {D{\"o}ring}, \citenamefont {Mussler},
  \citenamefont {Demarina}, \citenamefont {Luysberg}, \citenamefont
  {Bihlmayer}, \citenamefont {Sch{\"a}pers}, \citenamefont {Plucinski},
  \citenamefont {Bl{\"u}gel}, \citenamefont {Morgenstern}, \citenamefont
  {Schneider},\ and\ \citenamefont {Gr{\"u}tzmacher}}]{Eschbach15}%
  \BibitemOpen
  \bibfield  {author} {\bibinfo {author} {\bibfnamefont {M.}~\bibnamefont
  {Eschbach}}, \bibinfo {author} {\bibfnamefont {E.}~\bibnamefont
  {M{\l}y{\'n}czak}}, \bibinfo {author} {\bibfnamefont {J.}~\bibnamefont
  {Kellner}}, \bibinfo {author} {\bibfnamefont {J.}~\bibnamefont {Kampmeier}},
  \bibinfo {author} {\bibfnamefont {M.}~\bibnamefont {Lanius}}, \bibinfo
  {author} {\bibfnamefont {E.}~\bibnamefont {Neumann}}, \bibinfo {author}
  {\bibfnamefont {C.}~\bibnamefont {Weyrich}}, \bibinfo {author} {\bibfnamefont
  {M.}~\bibnamefont {Gehlmann}}, \bibinfo {author} {\bibfnamefont
  {P.}~\bibnamefont {Gospodari{\v c}}}, \bibinfo {author} {\bibfnamefont
  {S.}~\bibnamefont {D{\"o}ring}}, \bibinfo {author} {\bibfnamefont
  {G.}~\bibnamefont {Mussler}}, \bibinfo {author} {\bibfnamefont
  {N.}~\bibnamefont {Demarina}}, \bibinfo {author} {\bibfnamefont
  {M.}~\bibnamefont {Luysberg}}, \bibinfo {author} {\bibfnamefont
  {G.}~\bibnamefont {Bihlmayer}}, \bibinfo {author} {\bibfnamefont
  {T.}~\bibnamefont {Sch{\"a}pers}}, \bibinfo {author} {\bibfnamefont
  {L.}~\bibnamefont {Plucinski}}, \bibinfo {author} {\bibfnamefont
  {S.}~\bibnamefont {Bl{\"u}gel}}, \bibinfo {author} {\bibfnamefont
  {M.}~\bibnamefont {Morgenstern}}, \bibinfo {author} {\bibfnamefont {C.~M.}\
  \bibnamefont {Schneider}}, \ and\ \bibinfo {author} {\bibfnamefont
  {D.}~\bibnamefont {Gr{\"u}tzmacher}},\ }\href
  {http://dx.doi.org/10.1038/ncomms9816} {\bibfield  {journal} {\bibinfo
  {journal} {Nature Communications}\ }\textbf {\bibinfo {volume} {6}},\
  \bibinfo {pages} {8816} (\bibinfo {year} {2015})}\BibitemShut {NoStop}%
\bibitem [{\citenamefont {Nguyen}\ \emph {et~al.}(2016)\citenamefont {Nguyen},
  \citenamefont {Backes}, \citenamefont {Singh}, \citenamefont {Mansell},
  \citenamefont {Barnes}, \citenamefont {Ritchie}, \citenamefont {Mussler},
  \citenamefont {Lnius}, \citenamefont {Gr{\"u}tzmacher},\ and\ \citenamefont
  {Narayan}}]{Nguyen16}%
  \BibitemOpen
  \bibfield  {author} {\bibinfo {author} {\bibfnamefont {T.-A.}\ \bibnamefont
  {Nguyen}}, \bibinfo {author} {\bibfnamefont {D.}~\bibnamefont {Backes}},
  \bibinfo {author} {\bibfnamefont {A.}~\bibnamefont {Singh}}, \bibinfo
  {author} {\bibfnamefont {R.}~\bibnamefont {Mansell}}, \bibinfo {author}
  {\bibfnamefont {C.}~\bibnamefont {Barnes}}, \bibinfo {author} {\bibfnamefont
  {D.~A.}\ \bibnamefont {Ritchie}}, \bibinfo {author} {\bibfnamefont
  {G.}~\bibnamefont {Mussler}}, \bibinfo {author} {\bibfnamefont
  {M.}~\bibnamefont {Lnius}}, \bibinfo {author} {\bibfnamefont
  {D.}~\bibnamefont {Gr{\"u}tzmacher}}, \ and\ \bibinfo {author} {\bibfnamefont
  {V.}~\bibnamefont {Narayan}},\ }\href
  {https://www.nature.com/articles/srep27716} {\bibfield  {journal} {\bibinfo
  {journal} {Scientific Reports}\ }\textbf {\bibinfo {volume} {6}},\ \bibinfo
  {pages} {27716} (\bibinfo {year} {2016})}\BibitemShut {NoStop}%
\bibitem [{\citenamefont {Backes}\ \emph {et~al.}(2017)\citenamefont {Backes},
  \citenamefont {Huang}, \citenamefont {Mansell}, \citenamefont {Lanius},
  \citenamefont {Kampmeier}, \citenamefont {Ritchie}, \citenamefont {Mussler},
  \citenamefont {Gumbs}, \citenamefont {Gr{\"u}tzmacher},\ and\ \citenamefont
  {Narayan}}]{Backes17}%
  \BibitemOpen
  \bibfield  {author} {\bibinfo {author} {\bibfnamefont {D.}~\bibnamefont
  {Backes}}, \bibinfo {author} {\bibfnamefont {D.}~\bibnamefont {Huang}},
  \bibinfo {author} {\bibfnamefont {R.}~\bibnamefont {Mansell}}, \bibinfo
  {author} {\bibfnamefont {M.}~\bibnamefont {Lanius}}, \bibinfo {author}
  {\bibfnamefont {J.}~\bibnamefont {Kampmeier}}, \bibinfo {author}
  {\bibfnamefont {D.}~\bibnamefont {Ritchie}}, \bibinfo {author} {\bibfnamefont
  {G.}~\bibnamefont {Mussler}}, \bibinfo {author} {\bibfnamefont
  {G.}~\bibnamefont {Gumbs}}, \bibinfo {author} {\bibfnamefont
  {D.}~\bibnamefont {Gr{\"u}tzmacher}}, \ and\ \bibinfo {author} {\bibfnamefont
  {V.}~\bibnamefont {Narayan}},\ }\href {\doibase 10.1103/PhysRevB.96.125125}
  {\bibfield  {journal} {\bibinfo  {journal} {Phys. Rev. B}\ }\textbf {\bibinfo
  {volume} {96}},\ \bibinfo {pages} {125125} (\bibinfo {year}
  {2017})}\BibitemShut {NoStop}%
\bibitem [{\citenamefont {Backes}\ \emph {et~al.}(2019)\citenamefont {Backes},
  \citenamefont {Huang}, \citenamefont {Mansell}, \citenamefont {Lanius},
  \citenamefont {Kampmeier}, \citenamefont {Ritchie}, \citenamefont {Mussler},
  \citenamefont {Gumbs}, \citenamefont {Gr\"utzmacher},\ and\ \citenamefont
  {Narayan}}]{Backes19}%
  \BibitemOpen
  \bibfield  {author} {\bibinfo {author} {\bibfnamefont {D.}~\bibnamefont
  {Backes}}, \bibinfo {author} {\bibfnamefont {D.}~\bibnamefont {Huang}},
  \bibinfo {author} {\bibfnamefont {R.}~\bibnamefont {Mansell}}, \bibinfo
  {author} {\bibfnamefont {M.}~\bibnamefont {Lanius}}, \bibinfo {author}
  {\bibfnamefont {J.}~\bibnamefont {Kampmeier}}, \bibinfo {author}
  {\bibfnamefont {D.}~\bibnamefont {Ritchie}}, \bibinfo {author} {\bibfnamefont
  {G.}~\bibnamefont {Mussler}}, \bibinfo {author} {\bibfnamefont
  {G.}~\bibnamefont {Gumbs}}, \bibinfo {author} {\bibfnamefont
  {D.}~\bibnamefont {Gr\"utzmacher}}, \ and\ \bibinfo {author} {\bibfnamefont
  {V.}~\bibnamefont {Narayan}},\ }\href {\doibase 10.1103/PhysRevB.99.125139}
  {\bibfield  {journal} {\bibinfo  {journal} {Phys. Rev. B}\ }\textbf {\bibinfo
  {volume} {99}},\ \bibinfo {pages} {125139} (\bibinfo {year}
  {2019})}\BibitemShut {NoStop}%
\bibitem [{\citenamefont {Picozzi}(2014)}]{Picozzi14}%
  \BibitemOpen
  \bibfield  {author} {\bibinfo {author} {\bibfnamefont {S.}~\bibnamefont
  {Picozzi}},\ }\href {\doibase 10.3389/fphy.2014.00010} {\bibfield  {journal}
  {\bibinfo  {journal} {Frontiers in Physics}\ }\textbf {\bibinfo {volume}
  {2}},\ \bibinfo {pages} {10} (\bibinfo {year} {2014})}\BibitemShut {NoStop}%
\bibitem [{\citenamefont {Marder}(2010)}]{Marder_2010}%
  \BibitemOpen
  \bibfield  {author} {\bibinfo {author} {\bibfnamefont {M.~P.}\ \bibnamefont
  {Marder}},\ }\href {\doibase 10.1002/9780470949955} {\  (\bibinfo {year}
  {2010}),\ 10.1002/9780470949955}\BibitemShut {NoStop}%
\bibitem [{\citenamefont {Liebmann}\ \emph {et~al.}(2016)\citenamefont
  {Liebmann}, \citenamefont {Rinaldi}, \citenamefont {{Di Sante}},
  \citenamefont {Kellner}, \citenamefont {Pauly}, \citenamefont {Wang},
  \citenamefont {Boschker}, \citenamefont {Giussani}, \citenamefont {Bertoli},
  \citenamefont {Cantoni}, \citenamefont {Baldrati}, \citenamefont {Asa},
  \citenamefont {Vobornik}, \citenamefont {Panaccione}, \citenamefont
  {Marchenko}, \citenamefont {S{\'a}nchez-Barriga}, \citenamefont {Rader},
  \citenamefont {Calarco}, \citenamefont {Picozzi}, \citenamefont {Bertacco},\
  and\ \citenamefont {Morgenstern}}]{Liebmann16}%
  \BibitemOpen
  \bibfield  {author} {\bibinfo {author} {\bibfnamefont {M.}~\bibnamefont
  {Liebmann}}, \bibinfo {author} {\bibfnamefont {C.}~\bibnamefont {Rinaldi}},
  \bibinfo {author} {\bibfnamefont {D.}~\bibnamefont {{Di Sante}}}, \bibinfo
  {author} {\bibfnamefont {J.}~\bibnamefont {Kellner}}, \bibinfo {author}
  {\bibfnamefont {C.}~\bibnamefont {Pauly}}, \bibinfo {author} {\bibfnamefont
  {R.~N.}\ \bibnamefont {Wang}}, \bibinfo {author} {\bibfnamefont {J.~E.}\
  \bibnamefont {Boschker}}, \bibinfo {author} {\bibfnamefont {A.}~\bibnamefont
  {Giussani}}, \bibinfo {author} {\bibfnamefont {S.}~\bibnamefont {Bertoli}},
  \bibinfo {author} {\bibfnamefont {M.}~\bibnamefont {Cantoni}}, \bibinfo
  {author} {\bibfnamefont {L.}~\bibnamefont {Baldrati}}, \bibinfo {author}
  {\bibfnamefont {M.}~\bibnamefont {Asa}}, \bibinfo {author} {\bibfnamefont
  {I.}~\bibnamefont {Vobornik}}, \bibinfo {author} {\bibfnamefont
  {G.}~\bibnamefont {Panaccione}}, \bibinfo {author} {\bibfnamefont
  {D.}~\bibnamefont {Marchenko}}, \bibinfo {author} {\bibfnamefont
  {J.}~\bibnamefont {S{\'a}nchez-Barriga}}, \bibinfo {author} {\bibfnamefont
  {O.}~\bibnamefont {Rader}}, \bibinfo {author} {\bibfnamefont
  {R.}~\bibnamefont {Calarco}}, \bibinfo {author} {\bibfnamefont
  {S.}~\bibnamefont {Picozzi}}, \bibinfo {author} {\bibfnamefont
  {R.}~\bibnamefont {Bertacco}}, \ and\ \bibinfo {author} {\bibfnamefont
  {M.}~\bibnamefont {Morgenstern}},\ }\href {\doibase 10.1002/adma.201503459}
  {\bibfield  {journal} {\bibinfo  {journal} {Advanced Materials}\ }\textbf
  {\bibinfo {volume} {28}},\ \bibinfo {pages} {560} (\bibinfo {year}
  {2016})}\BibitemShut {NoStop}%
\bibitem [{\citenamefont {Sarma}\ and\ \citenamefont {Hwang}(2015)}]{Sarma15}%
  \BibitemOpen
  \bibfield  {author} {\bibinfo {author} {\bibfnamefont {S.~D.}\ \bibnamefont
  {Sarma}}\ and\ \bibinfo {author} {\bibfnamefont {E.~H.}\ \bibnamefont
  {Hwang}},\ }\href {\doibase 10.1038/srep16655} {\bibfield  {journal}
  {\bibinfo  {journal} {Scientific Reports}\ }\textbf {\bibinfo {volume} {5}},\
  \bibinfo {pages} {16655} (\bibinfo {year} {2015})}\BibitemShut {NoStop}%
\bibitem [{\citenamefont {Smidman}\ \emph {et~al.}(2017)\citenamefont
  {Smidman}, \citenamefont {Salamon}, \citenamefont {Yuan},\ and\ \citenamefont
  {Agterberg}}]{Smidman17}%
  \BibitemOpen
  \bibfield  {author} {\bibinfo {author} {\bibfnamefont {M.}~\bibnamefont
  {Smidman}}, \bibinfo {author} {\bibfnamefont {M.~B.}\ \bibnamefont
  {Salamon}}, \bibinfo {author} {\bibfnamefont {H.~Q.}\ \bibnamefont {Yuan}}, \
  and\ \bibinfo {author} {\bibfnamefont {D.~F.}\ \bibnamefont {Agterberg}},\
  }\href {http://stacks.iop.org/0034-4885/80/i=3/a=036501} {\bibfield
  {journal} {\bibinfo  {journal} {Reports on Progress in Physics}\ }\textbf
  {\bibinfo {volume} {80}},\ \bibinfo {pages} {036501} (\bibinfo {year}
  {2017})}\BibitemShut {NoStop}%
\bibitem [{\citenamefont {Yafet}(1963)}]{Yafet63}%
  \BibitemOpen
  \bibfield  {author} {\bibinfo {author} {\bibfnamefont {Y.}~\bibnamefont
  {Yafet}},\ }\enquote {\bibinfo {title} {g factors and spin-lattice relaxation
  of conduction electrons},}\ in\ \href@noop {} {\emph {\bibinfo {booktitle}
  {Solid State Physics Vol.14}}},\ \bibinfo {editor} {edited by\ \bibinfo
  {editor} {\bibfnamefont {F.}~\bibnamefont {Seitz}}\ and\ \bibinfo {editor}
  {\bibfnamefont {D.}~\bibnamefont {Turnbull}}}\ (\bibinfo  {publisher}
  {Academic Press},\ \bibinfo {address} {New York, London},\ \bibinfo {year}
  {1963})\BibitemShut {NoStop}%
\bibitem [{\citenamefont {Zheng}\ and\ \citenamefont
  {Das~Sarma}(1996)}]{PhysRevB.53.9964}%
  \BibitemOpen
  \bibfield  {author} {\bibinfo {author} {\bibfnamefont {L.}~\bibnamefont
  {Zheng}}\ and\ \bibinfo {author} {\bibfnamefont {S.}~\bibnamefont
  {Das~Sarma}},\ }\href {\doibase 10.1103/PhysRevB.53.9964} {\bibfield
  {journal} {\bibinfo  {journal} {Phys. Rev. B}\ }\textbf {\bibinfo {volume}
  {53}},\ \bibinfo {pages} {9964} (\bibinfo {year} {1996})}\BibitemShut
  {NoStop}%
\bibitem [{\citenamefont {Tsu}\ \emph {et~al.}(1967)\citenamefont {Tsu},
  \citenamefont {Howard},\ and\ \citenamefont {Esaki}}]{THE67}%
  \BibitemOpen
  \bibfield  {author} {\bibinfo {author} {\bibfnamefont {R.}~\bibnamefont
  {Tsu}}, \bibinfo {author} {\bibfnamefont {W.}~\bibnamefont {Howard}}, \ and\
  \bibinfo {author} {\bibfnamefont {L.}~\bibnamefont {Esaki}},\ }\href
  {\doibase https://doi.org/10.1016/0038-1098(67)90511-X} {\bibfield  {journal}
  {\bibinfo  {journal} {Solid State Communications}\ }\textbf {\bibinfo
  {volume} {5}},\ \bibinfo {pages} {167 } (\bibinfo {year} {1967})}\BibitemShut
  {NoStop}%
\bibitem [{\citenamefont {Narayan}\ \emph {et~al.}(2016)\citenamefont
  {Narayan}, \citenamefont {Nguyen}, \citenamefont {Mansell}, \citenamefont
  {Ritchie},\ and\ \citenamefont {Mussler}}]{Narayan16}%
  \BibitemOpen
  \bibfield  {author} {\bibinfo {author} {\bibfnamefont {V.}~\bibnamefont
  {Narayan}}, \bibinfo {author} {\bibfnamefont {T.-A.}\ \bibnamefont {Nguyen}},
  \bibinfo {author} {\bibfnamefont {R.}~\bibnamefont {Mansell}}, \bibinfo
  {author} {\bibfnamefont {D.~A.}\ \bibnamefont {Ritchie}}, \ and\ \bibinfo
  {author} {\bibfnamefont {G.}~\bibnamefont {Mussler}},\ }\href
  {http://onlinelibrary.wiley.com/doi/10.1002/pssr.201510430/abstract}
  {\bibfield  {journal} {\bibinfo  {journal} {physica status solidi - Rapid
  Research Letters}\ }\textbf {\bibinfo {volume} {10}},\ \bibinfo {pages} {253}
  (\bibinfo {year} {2016})}\BibitemShut {NoStop}%
\bibitem [{\citenamefont {Karvonen}\ and\ \citenamefont
  {Maasilta}(2007)}]{PhysRevLett.99.145503}%
  \BibitemOpen
  \bibfield  {author} {\bibinfo {author} {\bibfnamefont {J.~T.}\ \bibnamefont
  {Karvonen}}\ and\ \bibinfo {author} {\bibfnamefont {I.~J.}\ \bibnamefont
  {Maasilta}},\ }\href {\doibase 10.1103/PhysRevLett.99.145503} {\bibfield
  {journal} {\bibinfo  {journal} {Phys. Rev. Lett.}\ }\textbf {\bibinfo
  {volume} {99}},\ \bibinfo {pages} {145503} (\bibinfo {year}
  {2007})}\BibitemShut {NoStop}%
\bibitem [{\citenamefont {Lawrence}\ and\ \citenamefont
  {Wilkins}(1973)}]{Lawrence73}%
  \BibitemOpen
  \bibfield  {author} {\bibinfo {author} {\bibfnamefont {W.~E.}\ \bibnamefont
  {Lawrence}}\ and\ \bibinfo {author} {\bibfnamefont {J.~W.}\ \bibnamefont
  {Wilkins}},\ }\href {\doibase 10.1103/PhysRevB.7.2317} {\bibfield  {journal}
  {\bibinfo  {journal} {Phys. Rev. B}\ }\textbf {\bibinfo {volume} {7}},\
  \bibinfo {pages} {2317} (\bibinfo {year} {1973})}\BibitemShut {NoStop}%
\bibitem [{\citenamefont {Baral}\ \emph {et~al.}(2016)\citenamefont {Baral},
  \citenamefont {Vollmar}, \citenamefont {Kaltenborn},\ and\ \citenamefont
  {Schneider}}]{Baral16}%
  \BibitemOpen
  \bibfield  {author} {\bibinfo {author} {\bibfnamefont {A.}~\bibnamefont
  {Baral}}, \bibinfo {author} {\bibfnamefont {S.}~\bibnamefont {Vollmar}},
  \bibinfo {author} {\bibfnamefont {S.}~\bibnamefont {Kaltenborn}}, \ and\
  \bibinfo {author} {\bibfnamefont {H.~C.}\ \bibnamefont {Schneider}},\ }\href
  {\doibase 10.1088/1367-2630/18/2/023012} {\bibfield  {journal} {\bibinfo
  {journal} {New Journal of Physics}\ }\textbf {\bibinfo {volume} {18}},\
  \bibinfo {pages} {023012} (\bibinfo {year} {2016})}\BibitemShut {NoStop}%
\bibitem [{\citenamefont {Elliott}(1954)}]{PhysRev.96.266}%
  \BibitemOpen
  \bibfield  {author} {\bibinfo {author} {\bibfnamefont {R.~J.}\ \bibnamefont
  {Elliott}},\ }\href {\doibase 10.1103/PhysRev.96.266} {\bibfield  {journal}
  {\bibinfo  {journal} {Phys. Rev.}\ }\textbf {\bibinfo {volume} {96}},\
  \bibinfo {pages} {266} (\bibinfo {year} {1954})}\BibitemShut {NoStop}%
\bibitem [{\citenamefont {Overhauser}(1953)}]{PhysRev.89.689}%
  \BibitemOpen
  \bibfield  {author} {\bibinfo {author} {\bibfnamefont {A.~W.}\ \bibnamefont
  {Overhauser}},\ }\href {\doibase 10.1103/PhysRev.89.689} {\bibfield
  {journal} {\bibinfo  {journal} {Phys. Rev.}\ }\textbf {\bibinfo {volume}
  {89}},\ \bibinfo {pages} {689} (\bibinfo {year} {1953})}\BibitemShut
  {NoStop}%
\bibitem [{Note1()}]{Note1}%
  \BibitemOpen
  \bibinfo {note} {Note that we refer to the symmetry within the 2D plane,
  while obviously the Rashba interaction breaks the inversion symmetry in the
  direction perpendicular to the plane}\BibitemShut {NoStop}%
\bibitem [{Note2()}]{Note2}%
  \BibitemOpen
  \bibinfo {note} {Check Ref.~\cite {Baral16} for a detailed discussion of the
  low-$q$ expansion of the Elliot-Yafet mechanism.}\BibitemShut {Stop}%
\end{thebibliography}%

\end{document}